\documentclass[12pt]{article}
\pdfoutput=1
\usepackage{jheppub}
\usepackage{amsmath}
\usepackage{amsfonts}
\usepackage{amssymb}
\usepackage{graphicx}

\usepackage[export]{adjustbox}
\newcommand{\bmat}{\left(\begin{array}}
\newcommand{\emat}{\end{array}\right)}

\def\yzero{\smash{\hbox{$y\kern-4pt\raise1pt\hbox{${}^\circ$}$}}}

\def\beq{\begin{equation}}
\def\eeq{\end{equation}}
\def\beqa{\begin{eqnarray}}
\def\eeqa{\end{eqnarray}}

\def\-{\hphantom{-}}

\def\s2{\frac{1}{\sqrt2}}

\def\beq{\begin{equation}}
\def\eeq{\end{equation}}
\def\beqa{\begin{eqnarray}}
\def\eeqa{\end{eqnarray}}

\def\IF{\relax{\rm I\kern-.18em F}}
\def\II{\relax{\rm I\kern-.18em I}}

\def\Dsl{\,\raise.15ex\hbox{/}\mkern-13.5mu D} 

\def\IS{{\bf {S}}}
\def\IR{{\bf {R}}}

\def\IX{{\bf {X}}}

\def\IT{{\bf {T}}}

\def\NN{{\cal {N}}}

\newcommand{\eq}[1]{(\ref{#1})}

\catcode`\@=11   
\newdimen\@rotdimen
\newbox\@rotbox  

\def\@vspec#1{\special{ps:#1}}
\def\@rotstart#1{\@vspec{gsave currentpoint currentpoint translate
   #1 neg exch neg exch translate}}
\def\@rotfinish{\@vspec{currentpoint grestore moveto}}
\def\@rotr#1{\@rotdimen=\ht#1\advance\@rotdimen by\dp#1%
   \hbox to\@rotdimen{\hskip\ht#1\vbox to\wd#1{\@rotstart{90 rotate}%
   \box#1\vss}\hss}\@rotfinish}
\def\@rotl#1{\@rotdimen=\ht#1\advance\@rotdimen by\dp#1%
   \hbox to\@rotdimen{\vbox to\wd#1{\vskip\wd#1\@rotstart{270 rotate}%
   \box#1\vss}\hss}\@rotfinish}%
\def\@rotu#1{\@rotdimen=\ht#1\advance\@rotdimen by\dp#1%
   \hbox to\wd#1{\hskip\wd#1\vbox to\@rotdimen{\vskip\@rotdimen
   \@rotstart{-1 dup scale}\box#1\vss}\hss}\@rotfinish}%
\def\@rotf#1{\hbox to\wd#1{\hskip\wd#1\@rotstart{-1 1 scale}%
   \box#1\hss}\@rotfinish}%
\def\rotate{\@ifnextchar[{\@rotate}{\@rotate[l]}}
\def\@rotate[#1]#2{\setbox\@rotbox=\hbox{#2}\@nameuse{@rot#1}\@rotbox}

\catcode`\@=12

\begin{document}

\makeatletter
\@addtoreset{equation}{section}
\makeatother
\renewcommand{\theequation}{\thesection.\arabic{equation}}
\pagestyle{empty}
\rightline{IFT-UAM/CSIC-22-31}
\vspace{1.2cm}
\begin{center}
\Large{\bf At the End of the World:\\
Local Dynamical Cobordism}
\\

\large{Roberta Angius, Jos\'e Calder\'on-Infante, Matilda Delgado,  \\Jes\'us Huertas, Angel M. Uranga\\[4mm]}
\footnotesize{Instituto de F\'{\i}sica Te\'orica IFT-UAM/CSIC,\\[-0.3em] 
C/ Nicol\'as Cabrera 13-15, 
Campus de Cantoblanco, 28049 Madrid, Spain}\\ 
\footnotesize{\href{roberta.angius@csic.es}{roberta.angius@csic.es}, \href{mailto:j.calderon.infante@csic.es}{j.calderon.infante@csic.es}, \href{mailto:matilda.delgado@uam.es}{matilda.delgado@uam.es},
 \href{mailto:j.huertas@csic.es}{j.huertas@csic.es},   \href{mailto:angel.uranga@csic.es}{angel.uranga@csic.es}}

\vspace*{3mm}

\small{\bf Abstract} \\
\end{center}
\begin{center}
\begin{minipage}[h]{\textwidth}
The Cobordism Conjecture states that any Quantum Gravity configuration admits, at topological level, a boundary ending spacetime. We study the dynamical realization of cobordism, as spacetime dependent solutions of Einstein gravity coupled to scalars containing such end-of-the-world ‘branes’. The latter appear in effective theory as a singularity at finite spacetime distance at which scalars go off to infinite field space distance. We provide a local description near the end-of-the-world branes, in which the solutions simplify dramatically and are characterized in terms of a critical exponent, which controls the asymptotic profiles of fields and the universal scaling relations among the spacetime distance to the singularity, the field space distance, and the spacetime curvature. The analysis does not rely on supersymmetry. We study many explicit examples of such Local Dynamical Cobordisms in string theory, including 10d massive IIA, the 10d non-supersymmetric $USp(32)$ theory, Bubbles of Nothing, 4d $\NN=1$ cosmic string solutions, the Klebanov-Strassler throat, D$p$-brane solutions, brane configurations related to the D1/D5 systems, and small black holes. Our framework encompasses diverse recent setups in which scalars diverge at the core of defects, by regarding them as suitable end-of-the-world branes. We explore the interplay of Local Dynamical Cobordisms with the Distance Conjecture and other swampland constraints.

\end{minipage}
\end{center}
\newpage
\setcounter{page}{1}
\pagestyle{plain}
\renewcommand{\thefootnote}{\arabic{footnote}}
\setcounter{footnote}{0}

\tableofcontents

\vspace*{1cm}

\newpage

\section{Introduction}

The Cobordism Conjecture \cite{McNamara:2019rup} states that in any consistent theory of quantum gravity all cobordism classes are trivial. In simple terms, it must admit at the topological level a configuration ending spacetime\footnote{Equivalently, any two quantum gravity theories admit, at the topological level, a domain wall connecting them. For this paper we will emphasize the formulation as in the main text.}. Such end-of-the-world configuration may correspond to a boundary (such as the 10d Horava-Witten boundary of 11d M-theory \cite{Horava:1995qa,Horava:1996ma}), a bubble of nothing in which some compactification space shrinks to zero size \cite{Witten:1981gj} (see \cite{Ooguri:2017njy,GarciaEtxebarria:2020xsr,Dibitetto:2020csn,Bomans:2021ara} for some recent works), or more exotic possibilities, and may possibly be dressed with charged objects, such as branes, orientifold planes or generalizations (dubbed I-folds in \cite{Montero:2020icj}). The cobordism conjecture, already at this topological level, has produced interesting results, see \cite{GarciaEtxebarria:2020xsr,Ooguri:2020sua,Montero:2020icj,Dierigl:2020lai,Hamada:2021bbz,Blumenhagen:2021nmi} for some references\footnote{Spacetimes with boundaries have also been considered in the holographic setup, see \cite{Takayanagi:2011zk,Aharony:2011yc,Assel:2011xz,Bachas:2017rch,Bachas:2018zmb,
Raamsdonk:2020tin,VanRaamsdonk:2021duo} for some recent approaches.\label{foot:etw}}.

An exploration of the Cobordism Conjecture beyond the topological level was undertaken in \cite{Buratti:2021yia,Buratti:2021fiv} via the study of spacetime varying solutions to the equations of motion in theories with dynamical tadpoles, namely, potentials which do not have a minimum and thus do not admit maximally symmetric solutions (see \cite{Dudas:2000ff,Blumenhagen:2000dc,Dudas:2002dg,Dudas:2004nd} for early work and \cite{Basile:2018irz,Antonelli:2019nar,Basile:2020xwi,Basile:2021mkd} for related recent developments, and \cite{Draper:2021ujg,Draper:2021qtc} for a complementary approach to cobordism solutions). In the solutions in \cite{Buratti:2021yia,Buratti:2021fiv}, which we refer to as Dynamical Cobordisms, the fields run along a spatial coordinate until the solution hits a singularity at finite distance in spacetime, which (once resolved in the full UV theory) ends spacetime. 

These solutions exhibit sharp features in the region near the singularity. For instance, the scalars go off to infinite distance in moduli (or field) space at the spacetime singularity. Moreover, in the effective field theory description, the field space distance $D$, the spacetime curvature $R$ and the spacetime distance $\Delta$ to the singularity are related by interesting scaling laws, namely (in Planck units)
\beqa
\Delta \sim e^{-\frac 12\delta D}, \quad |R|\sim e^{\delta D}
\label{universal-scalings}
\eeqa 
for suitable positive coefficient $\delta$.

The singularities in these solutions are resolved in the full UV description, in terms of the corresponding cobordism configuration. In string theory examples, the latter often admits a tractable microscopic description involving geometries closing-off spacetime, possibly dressed with defects, as explained above. In this spirit, they were dubbed `cobordism defects' or `walls of nothing' in \cite{Buratti:2021yia,Buratti:2021fiv}. In this work we will mainly focus on the effective field theory description, where they remain as singular sources, which we refer to as End-of-The-World (ETW) branes\footnote{This follows the nomenclature in some of the references in footnote \ref{foot:etw}.}.

The universal form of the scaling relations \eqref{universal-scalings} was found by inspecting several explicit examples, but it suggests that a simple universal local description near the ETW branes should be possible in the effective theory. In this paper we provide this local description by studying Dynamical Cobordisms near walls at which the scalars run off to infinite field space distance. In this local description, the solutions simplify dramatically and are characterized in terms of a critical exponent $\delta$, which controls the asymptotic profiles of fields and the scaling relations \eqref{universal-scalings} in a very direct way. The analysis does not rely on supersymmetry and can be applied to non-supersymmetric setups.

This provides a powerful universal framework to describe ETW branes within effective field theory, as we illustrate in many different examples. We exploit it to describe Dynamical Cobordisms in several 10d string theories, including non-supersymmetric cases. We also use it to characterize warped throats \cite{Klebanov:2000nc,Klebanov:2000hb} as Dynamical Cobordisms. We moreover show that the familiar 10d D$p$-brane supergravity solutions can be regarded as Dynamical Cobordisms of sphere compactifications with flux, and are described by our local analysis with the D-branes playing the role of ETW branes. Finally, we argue that 4d small black hole solutions (see \cite{Mandal:2000rp,Dabholkar:2012zz} for some reviews), including those of the recent work \cite{Hamada:2021yxy}, can be similarly regarded as Dynamical Cobordisms of $\IS^2$ compactifications with flux, with the small black hole core playing the role of ETW brane.

Our models provide setups in which scalars explore large field space distances in a dynamical setup (as pioneered in \cite{Buratti:2018xjt}, see also \cite{Lust:2019zwm}), in contrast with the alternative adiabatic approach. Hence our description of Local Dynamical Cobordisms is the natural arena for the dynamical realization of swampland proposals\footnote{See \cite{Buratti:2018xjt} for a related viewpoint.} dealing with infinity in scalar moduli/field space.

The paper is organized as follows. In Section \ref{sec:local} we present the general formalism for the local description of Dynamical Cobordisms. In section \ref{sec:formulas} we present the general equations of motion, and in section \ref{sec:localdescription} we apply them to describe the local dynamics near ETW branes, and derive the universal scaling relations. In Section \ref{sec:10d} we apply the local description to several 10d examples, including massive IIA theory in section \ref{sec:massiveIIA} and the non-supersymmetric $USp(32)$ theory of \cite{Sugimoto:1999tx} in section \ref{sec:sugimoto}. In Section \ref{sec:branes} we interpret D-brane supergravity solutions as Dynamical Cobordisms (section \ref{sec:compactification}) and express them as ETW branes in the local description (section \ref{sec:local-dbranes}). Similar ideas are applied in section \ref{sec:eft-strings} to the EFT string in 4d $\NN=1$ theories in \cite{Lanza:2021qsu}, and in section \ref{sec:ks} to the Klebanov-Strassler warped throat \cite{Klebanov:2000nc,Klebanov:2000hb}. In Section \ref{sec:smallBHS} we discuss small black holes as Dynamical Cobordisms. In section \ref{sec:D6D2-T4S2} we warm up by expressing the supergravity solution of D2/D6-branes on $\IT^4$ as a Dynamical Cobordism, and in section \ref{sec:D6D2-T4T2S2} we relate it to small black holes via a further $\IT^2$ compactification. In section \ref{sec:hmvv-smallBHs} we consider more general small black holes, such as those in \cite{Hamada:2021yxy}, and derive scaling relations despite the absence of a proper Einstein frame in 2d. In Section \ref{sec:surprise} we discuss the interplay of Swampland constraints with the results of our local description for the behaviour of several quantities near infinity in field space. In section \ref{sec:swampland} we consider the Distance Conjecture, the de Sitter conjecture and the Transplanckian Censorship Conjecture. In section \ref{sec:largeN} we discuss potentially large backreaction effects when the UV description of the ETW branes involve a large number of degrees of freedom, suggesting mechanisms to generate non-trivial minima near infinity in field space. In section \ref{sec:conclusions} we offer some final thoughts. In Appendix \ref{app:curvature} we generalize the ansatz in the main text to allow for non-zero constant curvature in the ETW brane worldvolume directions (section \ref{sec:general-curved}), and apply it to describe Witten's bubble of nothing as a 4d Dynamical Cobordism and provide its local description (section \ref{sec:bon}). In Appendix \ref{app:subleading} we discuss subleading corrections to the local description, specially relevant in cases where the leading contributions vanish.

\section{Local Dynamical Cobordisms}
\label{sec:local}

In this section we formulate our local effective description near End of The World (ETW) branes, in terms of gravity coupled to a scalar field. We would like to emphasize that we consider a general scalar potential, but remarkably derive non-trivial results for its asymptotic behaviour near infinity in field space. The key input is just that the dynamics should allow for the scalar to go off to infinity in field space in a finite spacetime distance. 

Interestingly, the scalar potential generically behaves as an exponential near infinity in moduli/field space, suggesting a first-principles derivation of the `empirical' evidence for such exponential potentials, coming from string theory  and other swampland considerations (see \cite{Brennan:2017rbf,Palti:2019pca,vanBeest:2021lhn} for reviews). In particular, exponential potentials and constraints on them have been discussed in \cite{Draper:2021qtc,Draper:2021ujg}, for the restricted case of bubbles of nothing (i.e. UV completed to a purely geometrical higher dimensional configuration, \`a la \cite{Witten:1981gj}). In contrast, our analysis holds for fully general ETW branes (and hence, allows for more general potentials, including cases without this asymptotic growth).  

We focus on the case of a single scalar; however, our discussion also applies to setups with several scalars, by simply combining them into one effective scalar encapsulating the dynamics of the solutions (as illustrated in several of our examples in later sections). 

\subsection{General ansatz}
\label{sec:formulas}

Consider $d$-dimensional Einstein gravity coupled to a real scalar\footnote{Even though our analysis holds for general potential, we often refer to the scalar as modulus, and its field space as moduli space.} field with a potential, 
\begin{equation}
	S = \int d^{d}x\, \sqrt{-g}\,  \left( \frac{1}{2}R - \frac{1}{2} \left( \partial \phi\right)^{2} - V(\phi) \right) \, , 
	\label{ddim-action}
\end{equation}
where we are taking $M_{Pl}=1$ units. We focus on $d>2$, and deal with the $d=2$ case in some explicit examples in section \ref{sec:smallBHS}.

ETW branes define boundaries of the $d$-dimensional theory, hence they are described as real codimension 1 solutions. We take the ansatz
\beqa \label{dw-ansatz}\begin{aligned}
	ds^{2} & = e^{-2\sigma(y)} ds_{d-1}^{2} + dy^{2} \, ,\\
    \phi & = \phi(y) \, ,
   \end{aligned} \eeqa
where $y$ parametrizes the coordinate transverse to the ETW brane.

We consider flat metric in the $(d-1)$-dimensional slices. The corresponding analysis for general non-zero constant curvature, carried out in the same spirit and leading to essentially similar results, is presented in Appendix \ref{app:curvature}.

The equations of motion are
\begin{align}
	&\phi^{\prime\prime} - (d-1)  \sigma^{\prime} \phi^{\prime} - \partial_{\phi}V = 0 \, , \\
    &\frac{1}{2}(d-1)(d-2) \sigma^{\prime\,2} + V - \frac{1}{2} \phi^{\prime\,2} = 0 \, , \\
    &(d-2) \sigma^{\prime\prime} - \phi^{\prime\,2} = 0 \, ,
\end{align}
where prime denotes derivative with respect to $y$. The first one is the equation of motion for the scalar; for the Einstein equations, they split into transverse and longitudinal components to the ETW brane, giving two independent equations, subsequently combined into the last two equations.

The analysis of these equations is more amenable in terms of a new quantity, the tunneling potential introduced in \cite{Espinosa:2018hue,Espinosa:2018voj} (see also \cite{Espinosa:2018szu,Espinosa:2019hbm,Espinosa:2020qtq,Espinosa:2020cgk,Espinosa:2021qeo,Espinosa:2021tgx})
\begin{equation} \label{eq:tunneling-potential}
	V_{t}(\phi) \equiv V(\phi) - \frac{1}{2} \phi^{\prime\,2} \, .
\end{equation}

Using it to eliminate the scalar from the eoms we get
\begin{align}
	&(d-1) \sqrt{2\left(V-V_{t}\right)} \, \sigma^{\prime} - \partial_{\phi}V_{t} = 0 \, , \\
    &\frac{1}{2}(d-1)(d-2) \sigma^{\prime\,2} + V_{t} = 0 \, , \label{eq:sigma} \\
    &(d-2) \sigma^{\prime\prime} - 2\left(V-V_{t}\right) = 0 \label{eq:check}\, .
\end{align}

Finally, combining the first two equations to eliminate $\sigma$ we get
\begin{equation} \label{eq:Vt}
	\frac 14 (d-2) \left(\partial_{\phi}V_{t}\right)^{2} + (d-1) \left(V-V_{t}\right) V_{t} = 0 \, .
\end{equation}
This is a $d$-dimensional generalization of a condition found in \cite{Espinosa:2020cgk} in the context of domain walls.

Now, given a potential $V(\phi)$, one can use this equation to solve for the tunneling potential $V_{t}(\phi)$, and then use \eqref{eq:tunneling-potential} and \eqref{eq:sigma} to solve for $\phi(y)$ and $\sigma(y)$ respectively. In addition, one should check that \eqref{eq:check} is also satisfied.

Before moving on, let us comment on the implications that these equations have for the signs of the relevant quantities. From equation \eqref{eq:sigma} we learn that $V_{t}\leq 0$. In addition, from \eqref{eq:tunneling-potential} we get that $V-V_{t}\geq0$. Notice that these two facts are consistent with equation \eqref{eq:Vt}. Finally, combining the last inequality with \eqref{eq:check} we learn that $\sigma^{\prime\prime}\geq0$. When solving our system of equations we will systematically pick signs so that these inequalities are satisfied.

A nice way of parametrizing the freedom of choosing the potential is by writing
\begin{equation}
	V(\phi) = a(\phi) V_{t}(\phi) \, ,
	\label{eq:enter-a}
\end{equation}
where we have to impose that $a(\phi)\leq 1$ for the reason explained above. Plugging this into \eqref{eq:Vt} one can easily get to the solution
\begin{equation}
	\log \left( \frac{V_t}{V_{t}^{0}}\right) = \pm 2 \sqrt{\frac{d-1}{d-2}} \int_{\phi_0}^{\phi} \sqrt{1-a(\tilde{\phi})} \,d\tilde{\phi} \, ,
	\label{eq:vt_solution}
\end{equation}
where we are taking $V_{t}^{0}\equiv V_{t}(\phi_{0})$ as boundary condition.

\subsection{Local description of End of The World branes}
\label{sec:localdescription}

As explained in the introduction, we are interested in solutions for which the scalar attains infinity in field space i.e. $\phi\to\pm\infty$ at a point at finite distance in spacetime, defining an ETW brane. Without loss of generality we take this boundary to be $y=0$, and the infinity in field space as $\phi\to \infty$. 

From \eqref{eq:vt_solution}, it is clear that the asymptotic behavior as $y\to 0$, $\phi\to \infty$ is controlled by the asymptotic profile of $a(\phi)$. We know from the previous section that $a(\phi)\leq 1$ and we restrict our analysis to the cases where $a(\phi)$ has a well-defined and constant limit $a < 1$ as $\phi\to \infty$ (we briefly remark on the behavior $a\to 1$ below \eqref{eq:sol-V}). Indeed, although one can cook up potentials realizing other possibilities, we have not encountered them in any of the string theory examples in later sections. We therefore ignore other possibilities in what follows, leaving for future work the question about the consistency of such behaviors from the viewpoint of UV completions.  Note that the constraint $a < 1$ includes $a=0$, which corresponds to solutions with potential negligible with respect to the kinetic energy for the scalar (at least asymptotically).

Taking constant $a$, \eqref{eq:vt_solution} gives
\begin{equation} \label{eq:sol-Vt}
	V_{t}(\phi) \simeq - c\, e^{\delta\, \phi} \, ,
\end{equation}
where $c>0$ is related to the boundary condition used before. As explained in Appendix \ref{app:subleading}, we also allow $c$ to hide some $\phi$-dependence, corresponding to subleading corrections. The leading behaviour is an exponential controlled by the critical exponent $\delta$, given by
\begin{equation} \label{eq:gamma}
	\delta = 2 \sqrt{\frac{d-1}{d-2}\left( 1-a\right)} \, .
\end{equation}
Here we choose the plus sign for $\delta$. As we will see later this will imply that ETW brane explores $\phi\to\infty$ as explained above.

The critical exponent $\delta$  controls the structure of the local solution, in particular the asymptotic profile of fields as $y\to 0$, and the scaling relations among different physical local quantities.

Recall that the freedom of choosing a potential is parametrized by $a$. It is then interesting to ask how the potential itself looks like when approaching the end of the world. Plugging \eqref{eq:sol-Vt} into \eqref{eq:enter-a} we find
\begin{equation} \label{eq:sol-V}
	V(\phi) \simeq - a\,c \, e^{\delta\, \phi} \, .
\end{equation}
Note that we get an exponential dependence, for any value of $a<1$. As a side-note, for $a=1$, the potential $V$ may take different forms e.g. power-like, growing strictly slower than exponentials.

Also notice that, since $c>0$, the sign of the potential is completely determined by that of $a$. Moreover, using the relation between $a$ and the critical exponent $\delta$ in \eqref{eq:gamma}, we can put bounds on the latter depending on the sign of the potential. Namely, for $V>0$ we must have $a<0$, which implies $\delta> 2\sqrt{\frac{d-1}{d-2}}$, while if $V<0$ then $0<a<1$, yielding $\delta< 2\sqrt{\frac{d-1}{d-2}}$. We thus neither have negative potentials whose exponential behaviour is arbitrarily strong, nor positive potentials whose exponential behaviour is arbitrarily mild. The explanation is that such exponentials would lead to $\phi'^2 \gg V$ as we approach the ETW brane, and therefore they correspond to the $a=0$ case of our analysis.

It is interesting that we have derived fairly generically an exponential shape of the potential near infinity in moduli space, from the requirement that the theory contains ETW branes, namely configurations reaching infinity in moduli space at finite spacetime distance. In section \ref{sec:swampland} we will study its interplay with a variety of swampland constraints on scalar potentials. We note however that theories with milder growth of the potential (most prominently, theories with vanishing potential and exact moduli spaces) are still included in the analysis, and correspond to $a(\phi)\to0$. The corresponding statement that $V\to 0$ in this case actually means that the theory can have any potential as long as it grows slower than $\phi^{\prime\,2}$.

From \eqref{eq:tunneling-potential} we can obtain the asymptotic profile of $\phi$ as $y\to 0$ 
\begin{equation} \label{eq:sol-phi-full}
	\phi(y) \simeq - \frac{2}{\delta} \log \left( \frac{\delta^2}{4} \sqrt{2c\,\frac{d-2}{d-1}} \, y \right) \, .
\end{equation}
Here we are ignoring an additive integration constant, irrelevant in the $\phi\to\infty$ limit. We have also fixed another integration constant by demanding that the function blows up for $y\to 0$. The leading term as $y\to 0$ is
\begin{equation} \label{eq:sol-phi-simple}
	\phi(y) \simeq - \frac{2}{\delta} \log y \, .
\end{equation}
Hence the scalar goes off to infinity as we approach the end of the world. This motivates the appearance of a lowered cutoff as we approach the wall, above which a more complete microscopic description simply resolves the singularity; this resonates with the swampland distance conjecture, as we discuss in section \ref{sec:swampland}.

Plugging \eqref{eq:sol-phi-full} into \eqref{eq:sigma} we can also solve for $\sigma(y)$. The final result is
\begin{equation}\label{eq:sol-sigma}
	\sigma(y) \simeq \pm \frac{4}{(d-2)\delta^{2}} \,\log y \, .
\end{equation}
Here we ignore an integration constant which can be reabsorbed by a change of coordinates. Note that, to comply with \eqref{eq:check}, we only need to pick the minus sign.

Furthermore, the $d$-dimensional scalar curvature is given by
\begin{equation} \label{eq:sol-R}
	R = (d-1) \left(2\sigma^{\prime\prime} - d \sigma^{\prime\,2}\right) \sim \frac{1}{y^2} \, .
\end{equation}
We thus recover that the curvature blows up as we approach the end of the world, leading to a naked singularity in the effective field theory description. 

Notice that we have ignored a prefactor that, interestingly, vanishes for the special case $\delta^2 =\frac{2d}{d-2}$. For that value one should consider the next-to-leading order term in the $y\to 0$ expansion. In what follows we ignore this case and keep the generic one.

Since the scalar $\phi$ is normalized canonically, the field space distance $D$ as $y\to 0$ is \eqref{eq:sol-phi-simple}. Also, the distance in spacetime to the singularity is given by $y$. Hence from \eqref{eq:sol-phi-simple} and \eqref{eq:sol-R} we obtain the universal relations
\beqa \label{eq:scalingrelations}
    \Delta\sim e^{-\frac{\delta}{2}\,D}\quad ,\quad |R| \sim e^{\delta\,D} \, .
\eeqa

The solutions provides a simple universal description of dynamical cobordism in terms of the effective field theory. The microscopic description of the cobordism defect is available only in the UV complete theory, and is thus model-dependent (but known in many cases, see our explicit examples in later sections). From our present perspective, the only microscopic information we need is the very existence of such defects, guaranteed by the swampland cobordism conjecture \cite{McNamara:2019rup}\footnote{To be more precise, there are theories in which the cobordism higher-form symmetry is gauged, rather than broken by the existence of the defects. In such cases, the gauging imposes the constraint that the total charge cancels in the configuration; our analysis applies to those cases as well, with the ETW brane corresponding to a mere ending of spacetime with no explicit charged defects, similar to a bubble of nothing, see Appendix \ref{sec:bon}.}. It is thus remarkable that, the simple requirement that scalars go to infinity at finite spacetime distance leads to a complete local description of the EFT behaviour near a dynamical cobordism. Moreover, it constrains the structure of the theory, in particular it naturally yields an exponential behavior of the scalar potential near infinity in field space.

The above local description can be used to prove a general relation, introduced in  \cite{Buratti:2021yia}, between the dynamical tadpole  (defined as the derivative of the potential $\mathcal{T}=\partial_\phi V(\phi)$) at a given point and the spacetime distance $\Delta$ to the ETW brane, which in our examples is given by
\begin{equation}
    \Delta \sim \left(\mathcal{T} \right)^{-\frac{1}{2}} \, .
\end{equation}
Indeed, using \eqref{eq:sol-V} and \eqref{eq:sol-phi-simple}, we obtain $\mathcal{T}$ evaluated at a point $y^{*}$:
\begin{equation}
    \mathcal{T}|_{y=y^{*}}= \partial_\phi V|_{y=y^{*}} = - a \;c \;\delta \; e^{ \delta \phi}|_{y=y^{*}}= - a\;  c \; \delta\; (y^{*})^{-2} \, ,
\end{equation}
$\Delta$ is constructed as the distance from a point $y^{*}$ to the singularity at $y=0$, we therefore have $\Delta = y^{*}$. We hence have a general relation\footnote{For the particular case of the warped throat in \ref{sec:ks} this corrects the statement in \cite{Buratti:2021yia}.}
\begin{equation}
    \Delta = \left(\frac{-\mathcal{T}}{ a\; c\; \delta}\right)^{-\frac{1}{2}} \sim \left(\mathcal{T}\right)^{-\frac{1}{2}} \, .
\end{equation}
This relation places a bound on the spacetime extent of a solution whose running is induced by a dynamical tadpole, as emphasized in \cite{Buratti:2021yia,Buratti:2021fiv}, due the dynamical appearance of an end of spacetime. We would nevertheless mention that there exist solutions with spacetime boundaries even in situations with no dynamical tadpole. The simplest example is Horava-Witten theory, which corresponds to M-theory on an interval with two boundaries. Even in our present context of scalars running off to infinity at finite spacetime distance, it is possible to find ETW branes in cases with vanishing potential $V=0$ (or asymptotically negligible potentials, $a=0$).

\section{Some 10d Examples}
\label{sec:10d}

In this section we consider examples of 10d theories with Dynamical Cobordism solutions in \cite{Buratti:2021yia,Buratti:2021fiv}, and use the above local description to easily derive their structure. The results nicely match the asymptotic behavior of the complete solutions in the literature.

\subsection{The 10d massive type IIA theory}
\label{sec:massiveIIA}

We consider the 10d massive type IIA theory. The effective action in the Einstein frame for the relevant fields is
\begin{equation}
S_{10,E} = \frac{1}{2 \kappa^2} \int d^{10}x \sqrt{-g} \left\lbrace R -  \left( \partial \phi \right)^2 - \frac{1}{2} e^{\frac{5}{2} \sqrt{2} \phi} F_0^2 - \frac{1}{2} e^{\frac{\sqrt{2}}{2} \phi} \vert F_4 \vert^2 \right\rbrace \, ,
\end{equation}
where $F_0$ denotes the Romans mass parameter. The $\sqrt{2}$ factors in the exponents ensure that the normalization of the scalar agrees with our conventions.

This theory has a potential %
\begin{equation}
V =  \frac{1}{2} e^{\frac{5}{\sqrt{2}}  \phi} F_0^2,
\label{potentialMIIA}
\end{equation}
hence it does not admit 10d maximally symmetric solutions.
On the other hand there are 9d Poincar\'e invariant (and in fact 1/2 supersymmetric) running solutions of the equations of motion in which the dilaton (and other fields) depend on a space coordinate, e.g. $x^9$. The metric and dilaton profile read
\begin{equation}
\begin{split}
& ds_{10}^2= Z\left(x^9\right)^{1/12} \eta_{\mu \nu} dx^{\mu} dx^{\nu} \, , \\
& e^{\sqrt{2} \phi} = Z \left(x^9\right)^{-5/6} \, , \\
\end{split}
\label{massive-full-solution}
\end{equation}
where the coordinate function is $Z\left(x^9\right) = -F_0 x^9$.  This solution hits a singularity at $x^9 =0$, which was proposed to correspond to an end of the world brane in \cite{Buratti:2021yia,Buratti:2021fiv}. In the microscopic theory, it corresponds to an O8-plane (possibly with D8-branes), as in one of the boundaries of the interval of type I' theory \cite{Polchinski:1995df}. 

In the following we show how the local structure of the Dynamical Cobordism can be obtained from the analysis in the previous section.

The only input of the local analysis is the potential \eqref{potentialMIIA}. Matching it with the local analysis expression \eqref{eq:sol-V}, we obtain the following values for $\delta$ and, using \eqref{eq:gamma} for $a$:
\begin{equation}
\delta= \frac{5}{ \sqrt{2}} \quad , \quad  a= - \frac{16}{9}.
\end{equation} 
Plugging this into \eqref{eq:sol-phi-simple} we obtain the dilaton profile
\begin{equation}
\phi \simeq - \frac{2\sqrt{2}}{5 } \log y.
\label{massive-phi}
\end{equation}
We can now obtain the profile for $\sigma$ \eqref{eq:sol-sigma}
\beqa
\sigma \simeq -\frac{1}{25}\log y \ ,
\label{massive-sigma}
\eeqa
which determines the metric via \eqref{dw-ansatz}. As usual, the local description predicts the scalings
\begin{equation}
    \Delta \sim e^{- \frac{5}{2 \sqrt{2}}\, D} \quad , \quad |R| \sim e^{\frac{5}{\sqrt{2}}\, D} \, .
\end{equation}
These results from the local analysis are in agreement with the scaling relations obtained in the paper \cite{Buratti:2021yia} from the complete solution. In fact, this can be done very easily from \eqref{massive-full-solution}, by a change of coordinates
\begin{equation}
y = \int_0^{x^9} \left( - F_0 \tilde{x}^9 \right)^{1/24} d \tilde{x}^9,
\end{equation}
in terms of which the solution acquires the form of \eq{dw-ansatz}
\begin{equation}
\begin{split}
& ds_{10}^2 = \left[ \frac{25}{24} \left(- F_0 \right) y \right]^{2/25} ds_{9}^{2} +dy^2 \, , \nonumber\\
& e^{\sqrt{2} \phi} = \left[ \frac{25}{24} \left(- F_0 \right) y \right]^{-4/5} \, . \\
\end{split}
\end{equation}
This indeed corresponds to profiles for $\sigma$ (via \eqref{dw-ansatz}) and $\phi$ in agreement with \eqref{massive-sigma} and \eqref{massive-phi} respectively. 

\subsection{The 10d non-supersymmetric $USp(32)$ string}
\label{sec:sugimoto}

Let us consider a second example in the same spirit, but in the absence of supersymmetry. 
We consider the 10d non-supersymmetric $USp(32)$ theory, built in \cite{Sugimoto:1999tx} as a type IIB orientifold with a positively charged O9-plane and 32 anti-D9-branes. The 10d Einstein frame action for the relevant fields is 
\beqa
S_E\,=\, \frac{1}{2\kappa^2}\int d^{10}x \sqrt{-G}\,\left\lbrace\, R-(\partial\phi)^2\,\right\rbrace \,-\, T_9^E\int d^{10}x \sqrt{-G}\, 64\, e^{\frac{3}{\sqrt{2}}\phi}\, .
\label{usp32-action}
\eeqa
We have introduced factors of $\sqrt{2}$ relative to the conventions in \cite{Sugimoto:1999tx}, to normalize the scalar as in previous sections.

This theory has a dilaton tadpole, due to the uncanceled NSNS tadpoles, and hence does not admit maximally symmetric 10d solution. On the other hand, there are 9d Poincar\'e invariant running solutions of its equations of motion \cite{Dudas:2000ff}, given by
\beqa
ds_E^{\,2}&=& |\sqrt{\alpha_E}r|^{\frac 19}\, e^{-\frac{\alpha_E r^2}8} \eta_{\mu\nu} dx^\mu dx^\nu +  |\sqrt{\alpha_E}r|^{-1} e^{-\frac{3\phi_0}{\sqrt{2}}} e^{-\frac{9\alpha_E r^2}{8}}\, dr^2\,,\nonumber\\
\phi &=& \frac 3{4\sqrt{2}} \alpha_E r^2\,+\,\frac {\sqrt{2}}3\log|\sqrt{\alpha_E}r|\,+\,\phi_0 \, ,
\label{eoms-DM}
\eeqa
where $\alpha_E=64\kappa^2T_9^E$, and $\phi_0$ is a reference value for the dilaton. The coordinate $r$ was denoted by $y$ in \cite{Dudas:2000ff} but here, we preserve $y$ for the coordinate of the local analysis near end of the world branes.

The solution hits two singularities, at $r \to 0$ and at $r \to + \infty$, which are at finite spacetime distance, yet the scalar attains infinity in fields space ($\phi\to -\infty$ at $r\to0$, and $\phi \to \infty$ at $r\to \infty$, respectively). As discussed in \cite{Buratti:2021yia,Buratti:2021fiv}, it thus describes a Dynamical Cobordism with two end of the world branes. The existence of two boundaries, and hence a finite size spacetime coordinate, arises in this example, but is not a general feature of running solutions, as we have seen in previous sections. It would be interesting to understand a general criterion discriminating between the two possibilities, but we leave this question for future work. In any event, even in setups with two boundaries, our local analysis applies to each of them individually, as we discuss next. Indeed, let us now exploit the local analysis to display the scalings near these walls, with the scalar potential in \eqref{usp32-action} as sole input.

\subsubsection{$r\to 0$}

From equation \eqref{eoms-DM}, we see that $r\to 0$ corresponds to the limit $\phi\to-\infty$. The potential in \eqref{usp32-action} vanishes in that limit. As a consequence, we have an ETW brane in which the potential becomes negligible, i.e., the critical exponents for the local model are
\begin{equation}
\delta= \frac{3}{ \sqrt{2}} \quad , \quad  a= 0.
\label{massive-delta-a}
\end{equation} 

The local analysis then leads to the dilaton and radion profiles
\beqa
\phi\simeq \frac{2\sqrt{2}}3\log y\quad ,\quad
\sigma \simeq -\frac 19 \log y \, .
\label{sugi-sigma-phi}
\eeqa
Note that we have chosen the sign of $\phi\to -\infty$ as $y\to 0$.

These results allow to obtain the universal scalings for the curvature and spacetime distance with the field space distance \eqref{eq:scalingrelations}, namely
\beqa
\Delta\sim e^{-\frac{3}{2\sqrt{2}}D}\quad ,\quad |R|\sim e^{\frac{3}{\sqrt{2}}D} \, .
\label{scalings-usp}
\eeqa

It is easy to check that the above profiles and scaling reproduce the behaviour of the complete solution \eqref{eoms-DM}. This can be shown by the following coordinate change to bring it into the ansatz \eqref{dw-ansatz}:
\begin{equation}\label{y-r-DM}
    y=\int \sqrt{|\sqrt{\alpha_E}r|^{-1} e^{-\frac{3\phi_0}{\sqrt{2}}} e^{-\frac{9\alpha_E r^2}{8}}} dr \sim  \left[ \Gamma\left(\frac{1}{4},\frac{9\alpha_E}{16}r^2\right) - \Gamma\left(\frac{1}{4},0\right) \right] \sim \sqrt{r}  \, .
\end{equation}
In the last step we have taken the leading behaviour as $r\to 0$. By also taking the leading behaviour in \eqref{eoms-DM}, plugging in $y$, and reading off $\sigma$ as it appears in \eqref{dw-ansatz} we finally recover the profiles predicted by the local analysis in \eqref{sugi-sigma-phi}.

\subsubsection{$r\to \infty$}

This should be described by a local model where $\phi\to +\infty$ at $y\to 0$, i.e. the origin of a new local coordinate (which corresponds to $r\to \infty$). In this case the potential in \eqref{usp32-action} is blowing up, hence via \eqref{eq:sol-V} and \eqref{eq:gamma}, we get $\delta=3/\sqrt{2}$, $a=0$, just as in \eqref{massive-delta-a}. The result $a=0$ may seem puzzling, since from \eqref{eq:sol-V} this would seem to imply $V\to0$. However, one should recall that in the local description $a=0$ simply means that $V\ll \phi'^2$. Indeed, it may happen that $c$ blows up as $\phi\to\infty$ in such a way that it compensates having $a\to 0$ in this same limit. We will explicitly check this later on.

The dilaton and radion profiles read
\beqa\label{eq:DM:prediction2}
\phi\simeq -\frac{2\sqrt{2}}3\log y\quad ,\quad
\sigma\simeq -\frac 19 \log y \, .
\label{sugi-sigma-phi-2}
\eeqa
The dilaton sign differs from \eqref{sugi-sigma-phi} in order to have $\phi\to +\infty$ as $y\to 0$. We also recover the scalings for $\Delta$ and $R$ with $D$, which are again given by \eqref{scalings-usp}.

Let us now show that the above local model indeed reproduces the $r\to\infty$ regime of \eqref{eoms-DM}. 
The required change of variables is now
\begin{equation}
    y = \int_r^{\infty} \vert \sqrt{\alpha_E} \tilde{r} \vert^{- 1/2} e^{- \frac{3}{4} \phi_0} e^{- \frac{9\alpha_E r^2}{16}} d \tilde{r} \sim \Gamma \left( \frac{1}{4}, \frac{9\alpha_E}{16}r^2 \right) \sim r^{-\frac{3}{2}} e^{-\frac{9\alpha_E}{16}r^2} \, .
\end{equation}
The integration limits are chosen so that the finite distance singularity at $r \to \infty$ is located at the origin for the new coordinate. In the last step we have taken the leading behaviour of the Gamma function as $r\to \infty$.

Taking the logarithm of this expression and keeping the leading behaviour we get
\begin{equation}
    \log y \simeq - \frac{9\alpha_E}{16} r^2 \, .
\end{equation}
Finally, by also taking the leading behaviour in \eqref{eoms-DM}, reading off $\sigma$ as it appears in \eqref{dw-ansatz} and plugging in our previous expression for $y$, we recover the profiles anticipated by the local analysis in \eqref{eq:DM:prediction2}.

Let us now come back to the issue of having $a=0$ while not having vanishing potential. First, let us check that indeed $\phi'^2/V\to \infty$ as we approach the ETW brane. We can compute it, with no approximations, as
\beqa
    \frac{\phi'^2}{V} \sim \left( \frac{3\alpha_E}{2\sqrt{2}} r + \frac{\sqrt{2}}{3} \frac{1}{r} \right)^{2} \, ,
\eeqa
where we are ignoring irrelevant numerical prefactors. Importantly, for this computation one has to remember that $\phi'$ is the derivative with respect to $y$, not with respect to $r$. As advanced, we find that this blows up to infinity in both $r\to0$ and $r\to\infty$ limits.

Moreover, using this result one can compute the tunneling potential as $\phi\to\infty$ as
\begin{equation} \label{eq:DM-Vt}
    V_t \simeq \frac{\phi'^2}{2} \sim r^2 V \sim \phi\, e^{\frac{3}{\sqrt{2}}\phi} \sim e^{\frac{3}{\sqrt{2}}\phi + \log \phi }\, ,
\end{equation}
where we have plugged in the value of $V$ from \eqref{usp32-action} and $r^2\sim \phi$ from the $r\to\infty$ limit of $\phi(r)$ in \eqref{eoms-DM}. As advertised, we find a case in which the coefficient $c$ in \eqref{eq:sol-Vt} blows up as we approach the wall of nothing. This is consistent with our local analysis because, as we see in the last equality, $c$ does not blow up faster than the exponential, i.e., it gives subleading corrections to $\log V_t$ (see Appendix \ref{app:subleading} for more details).

\section{Branes as cobordism defects}
\label{sec:branes}

The local analysis of Section \ref{sec:local} provides a general framework to describe effective ETW branes, encapsulating Dynamical Cobordisms of the underlying theory. An interesting observation is that, in compactified theories with fluxes, the cobordism requires the introduction of charged objects. Namely, those required to break the corresponding cobordism charge to avoid a global symmetry, which should be absent in Quantum Gravity. A typical example is the introduction of NS5- and D-branes in bubbles of nothing in compactifications with NSNS and RR fluxes (see \cite{GarciaEtxebarria:2020xsr} for a recent discussion on bubbles of nothing).

Therefore it is interesting to explore the description of such objects in the local picture of  section \ref{sec:local}. As a simple illustrative setup, in this section we describe the geometry around a stack of D$p$-branes in the language of the local analysis of section \ref{sec:local}. In local terms, it corresponds to regarding the D$p$-brane supergravity solution as a compactification of the 10d theory on $\IS^{8-p}$ with flux, yielding a $d=(p+2)$-dimensional running solution along one of the coordinates (morally the radial coordinate), which has finite extent and end on an effective ETW brane. The microscopic description of the latter is actually given by the D$p$-brane in the UV.

The above idea generalizes the description in \cite{Buratti:2021fiv} of the EFT strings solutions in \cite{Lanza:2021qsu} as cobordism defects of $\IS^1$ compactifications of the underlying 4d $\NN=1$ theory with axion flux along the $\IS^1$.

We note that the compactification of the 10d theory on the $\IS^{8-p}$ around a D$p$-brane actually corresponds to a truncation onto the $SO(9-p)$-invariant sector. Sphere truncations have long been studied in the literature, in particular in the holographic context, see \cite{Itzhaki:1998dd} for a discussion for D$p$-brane solutions.
However, in our context we should regard the sphere truncation as a fair local description of Dynamical Cobordisms in actual compactifications, including those with scale separation, allowing for a more physical notion of lower-dimensional effective theory. Our local analysis should be regarded as part of the latter. This is depicted in Figure \ref{fig:brane}, and is illustrated quantitatively in a similar example for Witten's bubble of nothing in appendix \ref{sec:bon}.

\begin{figure}[htb]
\begin{center}
\includegraphics[scale=.4]{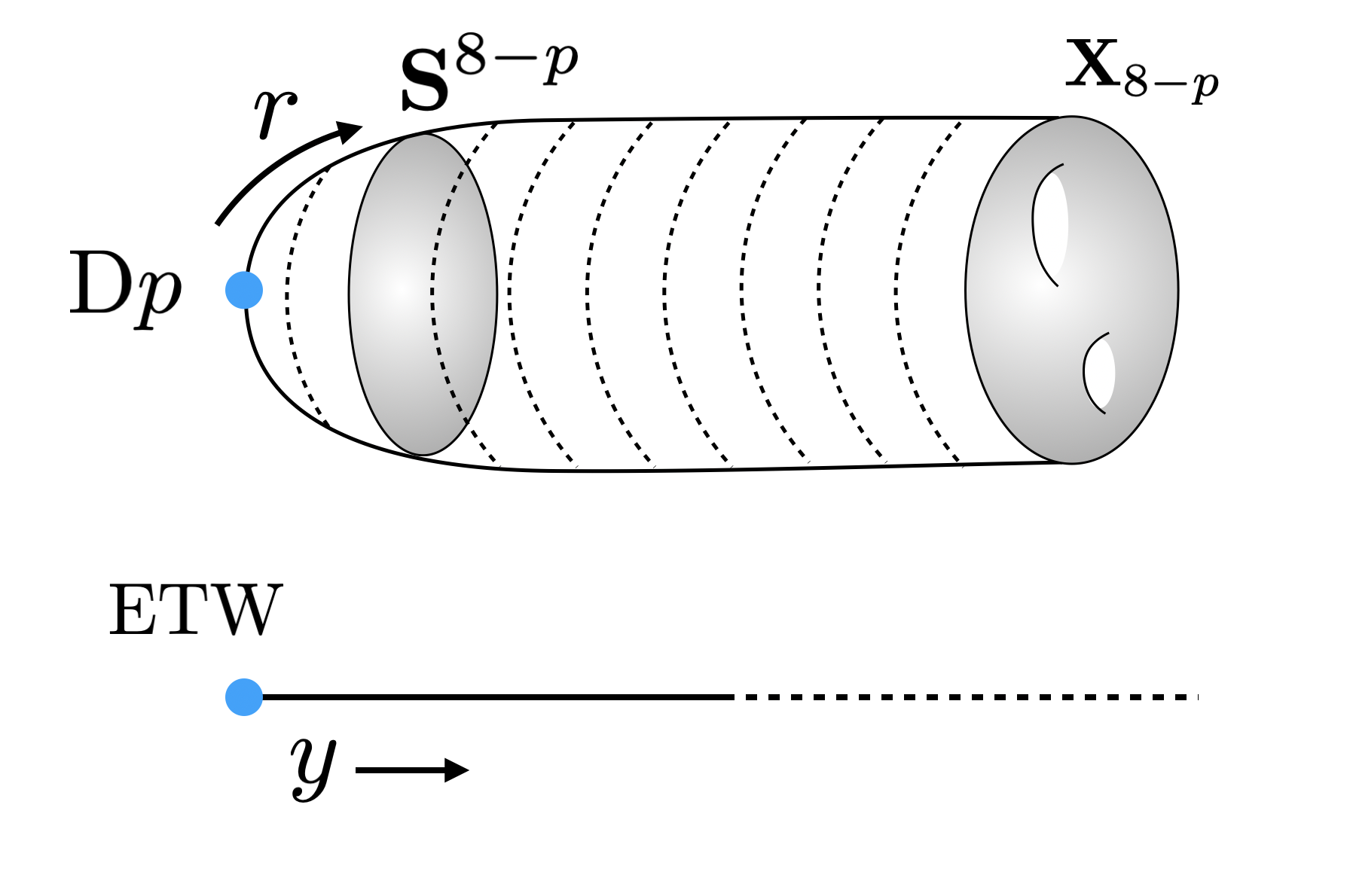}
\caption{\small D$p$-branes as cobordism defect in theories with $(8-p)$ compact dimensions from the higher and lower dimensional perspective. Our local $(p+2)$-dimensional description of the $\IS^{8-p}$-truncation corresponds to the local structure of a Dynamical Cobordism of a more general compactification on $\IX_{8-p}$.}
\label{fig:brane}
\end{center}
\end{figure}

Finally, although we phrase our discussion in terms of D$p$-branes, notice that other string theory branes admit similar analysis; in fact, the NS5-brane is essentially the same as the D5-brane, since we are working in the Einstein frame, in which S-duality acts manifestly.

\subsection{Compactification to a running solution}
\label{sec:compactification}

Let us begin with a precise description of the general procedure of compactifying a codimension $(n+1)$ brane-like solution in $d+n$ dimensions down to a running solution (codimension 1) in $d$ dimensions. In next sections, we will apply this reasoning to the D$p$-branes as cobordism defects of $\IS^{8-p}$ compactifications.

Take the general metric of a codimension $n$ object in $d+n$-dimensions: 
\begin{equation}\label{10dansatz}
    ds^2= e^{-2 \mu(r)}ds^2 _ {d-1} + e^{2 \nu(r)} (dr^2 + r^2 d\Omega^2_ n) \, .
\end{equation}
The directions in $ds^2_{d-1}$ span the worldvolume of the object, while we have split the transverse directions into radial and angular ones. 

We want to perform an $\IS^n$ truncation to look at this solution from the $d$-dimensional perspective. We thus take the compactification ansatz
\begin{equation}\label{eq:ansatz}
    ds^2 = e^{-2 \alpha \omega(r)} ds^2_{d} + e^{2 \beta \omega(r)}r_0^2 ds^2_n \, ,
\end{equation}
where $r_0$ is a reference scale. By requiring that the $d$-dimensional action is in the Einstein frame and has canonically normalized kinetic term for the radion $\omega$ we get the following constraints for $\alpha$ and $\beta$:
\begin{equation}\label{EinsteinFrame}
    \gamma \equiv \frac{\alpha}{\beta} = \frac{n}{d-2} \;\;\;\;\;\; \beta^2 = \frac{d-2}{n(d+n-2)} \, .
\end{equation}
The first one implements the Einstein frame requirement, while in the second one we already apply both conditions. Note that for $d=2$ we recover the familiar statement that there is no Einstein gravity in 2 dimensions. We will deal with reductions to 2d in section \ref{sec:smallBHS}, and consider $d>2$ in what follows.
 
By matching the compactification ansatz \eqref{eq:ansatz} with the metric in \eqref{10dansatz} we obtain the profile for the radion
\begin{equation}\label{radion}
    e^{2 \beta \omega(r)} = e^{2 \nu(r)} \left(\frac{r}{r_0}\right)^2 \, ,
\end{equation}
as well as the lower-dimensional metric 
\begin{equation}\label{compactansatz}
    ds^2_{d} = e^{2 \alpha \omega(r)}\left( e^{-2 \mu(r)}ds^2 _ {d-1} + e^{2 \nu(r)} dr^2\right) \, .
\end{equation}

In order to put solutions in the general form \eqref{dw-ansatz} used for the local description in section \ref{sec:local}, we introduce a new  coordinate
\begin{equation}\label{y-transformation}
   y = \int  e^{\alpha \omega(r)} e^{\nu(r)} dr \, ,
\end{equation}
in terms of which we can borrow the results \eqref{eq:sol-V}-\eqref{eq:scalingrelations} from the local description. 

From the viewpoint of the $d$-dimensional theory, there is a non-trivial potential arising from the curvature of the $\IS^n$, and possibly other sources (such as fluxes, etc). Generically the total potential does not have a minimum, hence the running solutions can be regarded as induced by a dynamical tadpole. Applying the results in \cite{Buratti:2021fiv,Buratti:2021yia}, the $d$-dimensional solution must describe a Dynamical Cobordism ending on an ETW brane, to which we can apply the local analysis in section \ref{sec:local}. 

Note that however there are cases with a non-trivial minimum. A prominent example  is the $\IS^5$ compactification with a large number $N$ of RR 5-form field string flux units (see \cite{Silverstein:2004id} for a discussion in similar terms). The minimum corresponds to a setup with no tadpole, and admits a maximally symmetric solution, namely the celebrated AdS$_5\times \IS^5$. Because of this, we will not consider the D3-branes in our discussion, and focus on genuinely running solutions.

\subsection{D-branes as Dynamical Cobordisms}
\label{sec:local-dbranes}

In this section we regard the 10d D$p$-brane solutions as $\IS^{8-p}$ compactifications and re-express them in terms of the local description of ETW branes of the $(p+2)$-dimensional theory of section \ref{sec:local}. Note that, in contrast with section \ref{sec:10d}, we do not intend to {\em derive} the local solutions from a $(p+2)$-dimensional scalar potential; rather we take the familiar 10d solutions and express their near brane asymptotics as local $(p+2)$-dimensional ETW brane solutions.

\medskip

Consider the D$p$-brane solution in the 10d Einstein frame, with $0\leq p \leq 8$. The 10d metric and dilaton profile take the form
\begin{align}\label{Dpbranesol}
    ds^2_{10} &= Z(r)^{\frac{p-7}{8}}\eta_{\mu\nu}dx^\mu dx^\nu +Z(r)^{\frac{p+1}{8}} (dr^2+r^2 d\Omega^2_{8-p})\, , \\
    \Phi &= \frac{(3-p)}{4\sqrt{2}}\log(Z(r)) \, ,
\end{align}
where the warp factor is given by the harmonic functions
\begin{align}
    Z(r) &= 1+\left(\frac{\rho}{r}\right)^{7-p} \text{ for } 0\leq p \leq 6 \, , \\
    Z(r) &= 1-\frac{N}{2\pi}\log\left(\frac{r}{\rho}\right) \text{ for } p=7 \, , \\
    Z(r) &= 1-\frac{|r|}{\rho} \text{ for } p=8 \, .
\end{align}
Here $\rho>0$ is a length scale. For the cases $p\neq7$ it depends on the number of D$p$-branes, $N$, while for $p=7$ this dependence does not enter in $\rho$ but has been made explicit in the solution. 

As we have explained, these formulas should be regarded as the local description near the D-branes in possibly more general compactifications, namely the above $Z(r)$ should be though of as local expansions around the D-brane location of the warp factor in more general compactification spaces, c.f. Figure \ref{fig:brane}.

We immediately see that for $p\neq 3$, the dilaton reaches infinite values near the point $r=0$, the core of the D$p$-brane. As explained above, we do not consider the case $p=3$, since it relates to AdS minimum of the theory. The solution does not run towards an ETW brane but towards a minimum in the potential. Similarly, for $p=8$, the dilaton reaches finite values at $r=0$. This fits with the identification of $D$8-branes as interpolating walls instead of walls of nothing in \cite{Buratti:2021fiv}. In the following we restrict to $p\neq 3$ and $1\leq p\leq 7$, the lower bound to avoid reduction to 2d (postponed until section \ref{sec:smallBHS}), and the upper bound to have non-trivial sphere compactification.

The D$p$-brane is a solution of the following generic type II theory with a dilaton and RR field:
\begin{equation}
    S_{10} \sim \frac{1}{2}\int d^{10}x \sqrt{-g_{10}}\left\{R_{10}-(\partial\Phi)^2 - \frac{1}{2 n!}e^{a \Phi}|F_n|^2\right\} \, .
\end{equation}
where $n=8-p$. This 10d theory does not have a scalar potential. However, once compactified on $\IS^{8-p}$ with $N$ units of $F_{8-p}$ flux, the curvature of the sphere as well as the flux itself will generate dynamical tadpoles for the ensuing radion and $(p+2)$-dimensional dilaton. Indeed, let us perform this compactification explicitly and show that we find ourselves in an end-of-the-world scenario, reproducing the associated scaling relations of \cite{Buratti:2021fiv}.\\

Taking a compactification ansatz of the form \eqref{eq:ansatz} we obtain the $d=(p+2)$-dimensional Einstein frame metric:
\begin{equation}\label{eq:lowdimmetric}
    ds^2_{d}= \left(\frac{r^2}{r_0^2} Z(r)^{\frac{p+1}{8}}\right)^{\frac{8-p}{p}} \left\{ Z(r)^{\frac{p-7}{8}}\eta_{\mu\nu}dx^\mu dx^\nu +Z(r)^{\frac{p+1}{8}} dr^2\right\} \, ,
\end{equation}
where the Greek indices correspond to directions along the world volume of the $p$-brane. The $(p+2)$-dimensional dilaton inherits the same profile as the original one and one obtains the radion's profile through matching: 
\begin{equation}\label{eq:RadionProf}
    e^{2\beta \omega(r)}= \frac{r^2}{r_0^2} Z(r)^{\frac{p+1}{8}} \, .
\end{equation}
The radion is canonically normalized if $\beta^2=\frac{p}{8(8-p)} \;$. 

The solution has a spacetime singularity at $r=0$, at which both the dilaton and radion blow up. We can now compute the relevant scaling quantities, namely the spacetime distance $\Delta_d$ to the singularity, the curvature scalar $|R_d|$ near the singularity, and the distance $D$ traversed in field space. For the former two we obtain:
\begin{equation}
   \Delta_{d}\sim     \begin{cases}
     r^{\frac{(p-3)^2}{2p}} \;\;\text{for $p\in[1,6]$ and $p\neq 3$} \, , \\ r^{8/7}  \;\;\text{for $p=7$}\, . 
\end{cases}
\end{equation}
\begin{equation}
   |R_{d}|\sim     \begin{cases}
     r^{-\frac{(p-3)^2}{p}}  \;\;\text{for $p\in[1,6]$ and $p\neq 3$} \, , \\ r^{-16/7}  \;\;\text{for $p=7$}\, . 
\end{cases}
\end{equation}
For the field space distance near the singularity, we obtain the following by plugging in the profiles of the radion \eqref{eq:RadionProf} and dilaton \eqref{Dpbranesol}: 
\begin{equation} \label{eq:DrDpbrane}
   D(r) = \int (d\omega^2+ d\Phi^2)^{1/2} dr  \simeq     \begin{cases}
     -\frac{|3-p|}{2}\sqrt{\frac{9-p}{p}}\log r  \;\;\text{for $p\in[1,6]$ and $p\neq 3$} \, , \\  - \frac{4}{\sqrt{14}} \log r   \;\;\text{for $p=7$}\, . 
\end{cases}
\end{equation}
The solution thus describes Dynamical Cobordisms with the following scaling relations: 
\begin{equation} \label{eq:scalingDp1to6}
     \Delta_{d}\sim  e^{\frac{|p-3|}{p} \sqrt{\frac{p}{9-p}} \,D} \, , \;\;\;\; |R_{d}|\sim e^{\frac{2|p-3|}{p} \sqrt{\frac{p}{9-p}} \,D} \;\;\text{for $p\in[1,6]$ and $p\neq 3$}\, , 
\end{equation}
and
\begin{equation}\label{eq:scalingDp7}
    \Delta_{9}\sim  e^{-\frac{2\sqrt{14}}{7} \,D} \, , \;\;\;\; |R_{9}| \sim e^{\frac{4\sqrt{14}}{7} \,D}   \;\;\text{for $p=7$}\, . 
\end{equation}

This shows that D$p$-brane are cobordism defects, which reduced on the surrounding $\IS^{8-p}$ can be described as ETW branes. In the following we describe their structure in terms of the local description of section \ref{sec:localdescription}. This will allow us a much simpler computation of the above scaling relations. 

The objective is to put the $d$-dimensional metric in domain-wall form \eqref{dw-ansatz}. In the notation of section \ref{sec:compactification}, one obtains: 
\begin{equation}\label{Sigma}
     \sigma( r)= - \alpha \omega(r)+ \mu(r)= -  \frac{8-p}{p}\log\left(Z(r)^{\frac{p+1}{16}} \left(\frac{r}{r_0}\right)\right) - \frac{p-7}{16}\log\left(Z(r)\right) \, .
\end{equation}
The new coordinate $y$ is obtained as
\begin{equation}\label{tildereq}
    y= \int^{ r} e^{\alpha \omega(r)} e^{\nu(r)}dr = \int^{ r}  Z( r)^{\frac{p+1}{2p}}  \left(\frac{r}{r_0}\right)^{\frac{8-p}{p} }dr \, .
\end{equation}
For a general D$p$-brane with $p\neq 3,7$, in the limit $r\rightarrow 0$ we have
\begin{equation}
    y=\int ^r \Big(\frac{\rho}{r}\Big)^{(7-p)\frac{p+1}{2p} } \Big(\frac{r}{r_0}\Big)^{\frac{8-p}{p}} dr \sim r^{\frac{(p-3)^2}{2p}} \, .
\end{equation}
Using equation \eqref{Sigma}, this yields
\begin{equation}
    \label{SIgmaDp}
     \sigma(r) \simeq \sigma\left( y^{\frac{2p}{(p-3)^2}}   \right) \simeq - \frac{(9-p)}{(p-3)^2}\log y \, .
\end{equation}
We may compare this to the profile for $\sigma$ put forward by the local description described in section \ref{sec:localdescription}:
\begin{equation}
	\sigma(y) \simeq - \frac{4}{p\;\delta^{2}} \,\log y \, .
\end{equation} We can thus extract the value of $\delta$ and, for completeness, that of $a$: 
\begin{equation}
    \delta^2 = \frac{4(p-3)^2}{p(9-p)} \, , \;\;\;\;\;\;\;\; 1- a=-\frac{(p-3)^2}{(p-9) (p+1)}\, .
\end{equation}
Thus, we have, from equation (\ref{eq:sol-phi-simple}): 
\begin{equation}\begin{aligned}
    D(y)&\simeq -\frac{2}{\delta}\log y \simeq -\frac{\vert p-3 \vert \sqrt{9-p}}{2 \sqrt{p\;  }} \log r \, .
\end{aligned}\end{equation}
We have thus recovered exactly the profile \eqref{eq:DrDpbrane}, without having to use the explicit scalar profile. From \eqref{eq:scalingrelations}, we also recover the scaling relations (\ref{eq:scalingDp1to6}), namely:
\begin{equation}
         \Delta_d =y \sim  e^{-\frac{|p-3|}{p} \sqrt{\frac{p }{9-p}} \,D} \, , \;\;\;\; |R_{d}|\sim e^{2\frac{|p-3|}{p} \sqrt{\frac{p }{9-p}} \,D} \, .
\end{equation}
Hence, in this case we have used the local description to recover the field-space distance and scaling relations near the singularity without knowing the full details of the $d$-dimensional theory. In fact, we can use the local description to derive the asymptotic behaviour of interesting $d$-dimensional quantities. For instance, the scalar potential scales near the singularity as \eqref{eq:sol-V}: 
\begin{equation}
    V(D)= - c\; \left(1-\frac{(p-3)^2}{(9-p) (p+1)}\right)\; e^{\frac{2(p-3)}{\sqrt{p(9-p)}}D} \, .
\end{equation}
This is a very interesting bottom-up approach. In the actual $d$-dimensional action, the potential would depend on the radion and dilaton with contributions  from the curvature of the sphere and the flux traversing it. However, the local description encapsulates only the dependence on the effective scalar dominating the field distance $D$ near the ETW brane, erasing any other irrelevant UV information. From the previous equation we find that the potential is negative as we approach the ETW brane (recall $c>0$). With the extra input that the curvature and the flux contributions to the potential are negative and positive respectively, the local description is then telling us that it is the curvature term the one that dominates in this limit.

For the D7-brane, the coordinate $y$ is given by
\begin{equation}
    y= \int^{ r} e^{\alpha \omega(r)} e^{\nu(r)}dr = \int^{ r} \Big(-\frac{N}{2\pi}\log\left(\frac{r}{\rho}\right) \Big)^{\frac{4}{7}}  \left(\frac{r}{r_0}\right)^{\frac{1}{7} }dr \sim  r^{\frac{8}{7}} \, ,
\end{equation}
where we have neglected the logarithmic contribution compared to the polynomial one. Similarly, we have:
\begin{equation}
    \label{SIgmaD7}
     \sigma(r) \simeq \sigma(y^{\frac{7}{8}}) \simeq -\alpha \omega(y^{\frac{7}{8}})  \simeq -\frac{1}{8}\log y \, .
\end{equation}
Hence, comparing this to equation \eqref{eq:sol-sigma}, we find:
\begin{equation}
    \delta^2 = \frac{32}{7}\, ,  \;\;\;\;\;\;\;\; a=0\,.
\end{equation} 
This means that the asymptotic potential vanishes, in the sense of $\phi'^2\gg V$. Plugging this value of $\delta^2$ into equation (\ref{eq:sol-phi-simple}) and \eqref{eq:scalingrelations}, we recover the same field space distance and scaling relations as in the computations of the previous section:
\begin{equation}
    D(y) \simeq -\sqrt{\frac{7}{8 }} \log y \simeq -\frac{4}{\sqrt{14 }}\log r \, , 
\end{equation}
\begin{equation}
         \Delta_9 =y \sim  e^{ -\sqrt{\frac{8 }{7}} \,D} \quad ,\quad |R_{9}|\sim e^{2 \sqrt{\frac{8 }{7}} \,D} \, .
\end{equation}

\subsection{Revisiting the EFT strings}
\label{sec:eft-strings}

In \cite{Lanza:2020qmt,Lanza:2021qsu} it was proposed that in 4d $\NN=1$ theories the limits in which saxionic scalars go to infinity in moduli space can be studied as radial flows in 4d supersymmetric EFT string solutions magnetically charged under the corresponding axionic partners. In \cite{Buratti:2021fiv} the result was recovered by considering running solution of the compactification of the theory to 3d with axion fluxes along the $\IS^1$: the solutions implement a Dynamical Cobordism ending spacetime along the running direction, and the EFT string arises as the cobordism defect required to get rid of the axion flux. In this section we revisit the analysis in \cite{Buratti:2021fiv} from the local description, with the EFT string becoming an ETW brane. As expected, the analysis is fairly similar to the 10d D7-brane example in the previous section; indeed, upon compactification of the 10d theory on a CY3, the wrapped D7-branes turns into the simplest avatar of the EFT strings in \cite{Lanza:2020qmt,Lanza:2021qsu}.

In the 4d EFT string solution \cite{Lanza:2020qmt,Lanza:2021qsu}, the profile for the scalars is given by
\begin{align}
    s(r) &= s_0 - \frac{q}{2\pi} \log\frac{r}{r_0}\, , \label{sol-s}\\
    a(\theta) &= a_0 + \frac{\theta}{2\pi} q \label{sol-a} \, .
\end{align}
In our 3d interpretation, equation \eqref{sol-a} describes the axionic flux over the $\IS^1$, and equation \eqref{sol-s} solves the dynamical tadpole for the saxion. 

The 4d metric takes the form 
\begin{equation} \label{eq:axionic-metric}
    ds_4 ^2= -dt^2 +dx^2 + e^{2D}d z d\bar z\, ,
\end{equation}
with $z=r e^{i\theta}$. The warp factor is given by
\begin{equation} \label{eq:4d-warping}
    2D=-K+K_0=\frac{2}{n^2}\log{\frac{s}{s_0}} \, ,
\end{equation}
where the K\"{a}hler potential is  $K=-\frac{2}{n^2}\log{s}$. This $D$ should not be confused with the field space distance, and we trust the reader to distinguish them by the context.

Matching the 4d metric \eqref{eq:axionic-metric} to the setup in section \ref{sec:compactification} with $n=1$, we obtain the 3d coordinate $y$:
\begin{equation}\begin{aligned}
    y& = \int^{r}  e^{\alpha \omega(r)} e^{\nu(r)}dr = \int ^r \left(1-\frac{q}{2\pi s_0}\log \frac{r}{r_0}\right)^{\frac{2}{n^2}}\frac{r}{r_0}dr\sim r^2 \, ,
\end{aligned}\end{equation}
where we have once more neglected the logarithm compared to the polynomial contribution. Then, we can put the 3d metric in the domain-wall form  \eqref{dw-ansatz}, in the $r\rightarrow 0$ limit, with:
\begin{equation}
    \sigma(y^{\frac{1}{2}})=- \gamma \beta \omega(y^{\frac{1}{2}}) \simeq - \log \Bigg(\left(1-\frac{q}{2\pi s_0}\log \frac{y^{\frac{1}{2}}}{r_0}\right)^{\frac{1}{n^2}}\frac{y^{\frac{1}{2}}}{r_0}\Bigg) \simeq -\frac{1}{2} \log y \, .
\end{equation}
Comparing this to \eqref{eq:sol-sigma}, we obtain
\begin{equation}
    \delta^2 = 8\quad,\quad a=0\;.
\end{equation}
We can use these parameters to recover the profiles and scaling of the local solution. For instance, we obtain that $\phi'^2\gg V$, as in the D7-brane case. We also obtain the field-space profile and scaling relations\footnote{This result corrects a factor of $\sqrt{2}$ arising from  $|D| \simeq |\sigma_p| \simeq -\sqrt{2}\log(r)$, which was missing in \cite{Buratti:2021fiv}.} from \eqref{eq:sol-phi-simple} and \eqref{eq:scalingrelations}:
\begin{equation}\begin{gathered}
D(y) \simeq -\sqrt{\frac{1}{2}} \log y \, ,
\end{gathered}\end{equation}
\begin{equation}\begin{gathered}
\Delta = y \sim e^{-\sqrt{2}\,D}\; , \;\;\;\;\;\;
|R|=e^{2\sqrt{2}\,D}\; .
 \end{gathered}\end{equation}

We thus find that the full  solution can be described  in terms of the local description, with the EFT string described in terms of an ETW brane. 

\subsection{The Klebanov-Strassler throat}
\label{sec:ks}

In the previous examples we have shown that D-branes can play the role of ETW branes in running solutions of compactifications with fluxes. We would like to mention, however, an alternative mechanisms in which Dynamical Cobordisms can get rid of fluxes in the compactification, namely when the running involves axion monodromy\footnote{For axion monodromy in inflation, see \cite{Silverstein:2008sg,Ibanez:2014kia,McAllister:2008hb,Kaloper:2008fb,Flauger:2009ab,Marchesano:2014mla,Hebecker:2014eua,Franco:2014hsa}}.
This is most clearly illustrated in the celebrated Klebanov-Strassler (KS) solution \cite{Klebanov:2000hb}, related to the compactification of type IIB theory on the 5d Sasaki-Einstein space $T^{1,1}$ with $N$ units of RR 5-form flux and $M$ units of RR 3-form flux on an $\IS^3\subset T^{1,1}$. 

As shown in \cite{Buratti:2021yia}, the KS solution can be regarded as a Dynamical Cobordism, in which the tip of the throat ends spacetime at finite spacetime distance in the radial direction, smoothing out (or UV completing) the singularity of the related Klebanov-Tseytlin (KT) solution \cite{Klebanov:2000nc}. In this section we show that the structure of the KT solution is indeed that of an ETW brane from the viewpoint of the 5d effective theory.

Consider the KT solution \cite{Klebanov:2000nc}, whose 10d Einstein frame metric reads:
\begin{equation} \label{eq:10d-metric}
	ds_{10}^{2} = h^{-1/2}(r) \eta_{\mu\nu} dx^{\mu}dx^{\nu} + h^{1/2}(r) \left( dr^2 + r^{2} ds_{T^{1,1}}^{2}\right) \, ,
\end{equation}
with
\begin{equation}
	h(r) = b_{0} + \frac{M^2 \log\left( r/r_{*}\right)}{4r^{4}} \, .
\end{equation}

The singularity is at $r_{s}$ such that $h(r_{s})=0$, signalling the location of the ETW brane. One can show that $\partial_rh\neq 0$ at $r=r_s$, hence we may expand this harmonic function near this point as
\begin{equation}
	h(r) \sim r-r_{s} \equiv \tilde{r} \, .
\end{equation}
We now take the compactification ansatz
\begin{equation}
	ds_{10}^{2} = L^{2} \left( e^{-5q} ds_{5}^{2} + e^{3q} ds_{T^{1,1}}^{2}\right) \,
\end{equation}
with $L$ an overall scale. Matching with \eqref{eq:10d-metric} we get the profile for the breathing mode
\begin{equation} \label{eq:q-profile}
	q(r) = \frac{1}{6} \log \left( \left(\frac{r}{L}\right)^{4} h(r) \right) \simeq \frac{1}{6} \log \tilde{r} \, ,
\end{equation}
where in the last equality we have taken the near ETW limit. We also get the 5d Einstein frame metric:
\begin{equation}
	L^{2} ds_{5}^{2} = \left( \left( \frac{r}{L}\right)^{2} h^{\frac{1}{2}}\right)^{\frac{5}{3}} \left( h^{-\frac{1}{2}} \eta_{\mu\nu} dx^{\mu}dx^{\nu} + h^{\frac{1}{2}} dr^{2} \right)  \, .
\end{equation}
From it we can derive the relation between $\tilde{r}$ and the radial coordinate $y$ in the local analysis, which is
\begin{equation} \label{eq:y-coordinate}
	\tilde{r} \sim y^{\frac{3}{5}} \, .
\end{equation}
Reading off the warp factor
\begin{equation}
	e^{-2\sigma} = \left( \left( \frac{r}{L}\right)^{2} h^{\frac{1}{2}}\right)^{\frac{5}{3}} h^{-\frac{1}{2}} \sim \tilde{r}^{\frac{1}{3}} \sim y^{\frac{1}{5}} \, ,
\end{equation}
we finally find
\begin{equation}
	\sigma(y) \simeq - \frac{1}{10} \log y \, .
\end{equation}
Hence, the 5d KT solution near the singularity fits with the form of an ETW brane in our local description with
\begin{equation} \label{eq:coeffs}
	\delta = \frac{2\sqrt{30}}{3} \quad , \quad a = -\frac{3}{2} \, .
\end{equation}
We can also check that the solution for the scalars also fits in the local model description. The NSNS axion is given by
\begin{equation}
	T(r) = \tilde{T} + M \log r \simeq T_{s} + \frac{M}{r_s} \tilde{r} \, ,
\end{equation}
again in the near ETW brane limit. 
Here $T_{s}=T(r_s)$, which we can keep arbitrary. The field space metric from the 5d action in \cite{Klebanov:2000nc} is given by
\begin{equation}
	dD^{2} = 30 (\partial q)^{2} + \frac{1}{2} g_{s}^{-1} e^{-6q} (\partial T)^{2} \, .
\end{equation}
Using the profiles for $q$ and $T$ in the ${\tilde r}\to 0$ limit, we have
\beqa 
	(\partial q)^{2} \simeq \frac{1}{36\tilde{r}^{2}} \quad \quad,\quad \quad
    e^{-6q}(\partial T)^{2} \simeq \left(\frac{M}{r_s}\right)^2 \,\frac{1}{\tilde{r}} \, .
\eeqa 
For $\tilde{r}\to0$, the breathing modes dominates the field space distance in field-space. Following \cite{Calderon-Infante:2020dhm}, it is then an asymptotically geodesic trajectory. This is in contrast with the $r\to\infty$ limit, for which the field-space trajectory was shown to be highly non-geodesic in \cite{Buratti:2018xjt}. Hence we have
\begin{equation} \label{eq:D}
	dD^{2} \simeq 30 (\partial q)^{2} \simeq \frac{5}{6} \tilde{r}^{-2} \, .
\end{equation}
Upon integration and using \eqref{eq:y-coordinate} we obtain
\begin{equation}
	D(y) \simeq - \frac{\sqrt{30}}{10} \log y \, .
\end{equation}
This again takes the form found in our local analysis, for the above coefficients \eqref{eq:coeffs}.

Finally, we also check that the 5d scalar potential from \cite{Klebanov:2000nc} scales as predicted by the local model. The complete potential is
\begin{equation}
	V(\phi) = - 5 e^{-8q} + \frac{1}{8} g_{s} M^{2} e^{-14q} + \frac{1}{8} (N+MT)^{2} e^{-20q} \, .
\end{equation}
Plugging in $T=T_{s}$ and $D\simeq-\sqrt{30}\,q$ as dictated by \eqref{eq:D}, we get
\begin{equation}
	V(D) = - 5 e^{\frac{4\sqrt{30}}{15}D} + \frac{1}{8} g_{s} M^{2} e^{\frac{7\sqrt{30}}{15}D} + \frac{1}{8} (N+MT_{s})^{2} e^{\frac{2\sqrt{30}}{3}D} \, .
\end{equation}
For $N+MT_{s}\neq 0$, we find that the last term dominates as $D\to\infty$. As predicted by our local analysis, it has an exponential behaviour with $D$ with the coefficient $\delta$ given in \eqref{eq:coeffs}. Moreover, as predicted by finding $a<0$, the coefficient in front of this exponential is positive.

We hope these examples suffice to convince the reader that the local description provides a simple and efficient framework to discuss the structure of Dynamical Cobordisms near the ETW brane.

\section{Small Black Holes as Dynamical Cobordisms}
\label{sec:smallBHS}

The analysis of the previous section for single-charge D-brane solutions can be similarly carried out for systems of multiple charges, namely combining D-branes of different dimensionalities. Such systems have been extensively employed in the construction and microscopic understanding of black holes, both with finite horizon, starting with \cite{Strominger:1996sh}, or with vanishing classical horizon area (small black holes) (see \cite{Mandal:2000rp,Dabholkar:2012zz} for some reviews). In this section we describe brane configurations, closely related to the celebrated D1/D5 system, leading to small black holes, and describe them as cobordism defects of suitable sphere compactifications of the underlying theory. The resulting dimensionally truncated theory corresponds to a 2d theory of gravity and an effective scalar (2d dilaton gravity), for which we find scaling relations analogous to the higher dimensional cases. This description relates the Dynamical Cobordisms to the realization of the Swampland Distance Conjecture in small black holes\footnote{For other approaches to Swampland constraints using (large and small) black holes, see e.g. \cite{Bonnefoy:2020fwt,Luben:2020wix,Cribiori:2022cho}.} in \cite{Hamada:2021yxy}.

\subsection{The D2/D6 system on $\IT^4$}
\label{sec:D6D2-T4S2}

We consider a configuration of D6- and D2-branes in the following (1/4 susy preserving) configuration
\begin{align}
    D6:& \;\; 0\;\;1\;\;2\;\;\times\;\;\times\;\;\times\;\;6\;\;7\;\;8\;\;9\\
    D2:& \;\; 0\;\;1\;\;2\;\;\times\;\;\times\;\;\times\;\;\times\;\;\times\;\;\times
\end{align}
where the numbers correspond to directions spanned by the brane worldvolumes and $\times$'s mark transverse directions. We consider all branes to coincide in the mutually transverse directions 345. We moreover smear the D2-branes in the direction 6789. Eventually these directions will be taken to be compact, so the smeared description is valid for small compactification size.\\

In the 10d Einstein frame the metric and dilaton profile are given by harmonic superposition (see \cite{Ortin:2015hya} for background)
\beqa
\label{10dsmeared}
   & ds^2 = Z_6(r)^{-\frac{1}{8}}Z_2(r)^{-\frac{5}{8}} \eta_{\mu\nu} dx ^\mu dx^\nu + Z_6(r)^{\frac{7}{8}}Z_2(r)^{\frac{3}{8}}  (\,dr^2+r^2 d\Omega^2_2\,) +  Z_6(r)^{-\frac{1}{8}}Z_2(r)^{\frac{3}{8}} dx^mdx^m&\nonumber , \\
&\Phi(r)= \frac{1}{2\sqrt{2}} \log\left(Z_6(r)^{-\frac{3}{2}} Z_2(r)^{\frac{1}{2}}\right) \, , \quad\quad\quad\quad\quad\quad\quad\quad &
\eeqa
where $r$ is the radial coordinate in 345, $d\Omega_2^2$ is the volume of a unit $\IS^2$ in this $\IR^3$, and $m=6,7,8,9$. The harmonic functions are
\beqa
Z_6(r) = 1+\frac{\rho_6}{r}\quad ,\quad
Z_2(r)= 1+ \frac{\rho_2 }{r} \, .
\eeqa

As announced, we now consider compactifying the directions $6789$ on a $\mathbf{T}^4$ (similar results hold for K3 compactification, as usual), with the compactification ansatz
\begin{equation}
    ds^2 = e^{-\frac{t}{\sqrt{2}}}ds^2_6 + e^{\frac{t}{\sqrt{2}}}ds^2_{T^4} \, .
\end{equation}
Matching this ansatz to (\ref{10dsmeared}), we obtain the canonically normalized radion
\begin{equation}
    t(r) = \sqrt{2} \log\left(Z_6(r)^{-\frac{1}{8}}Z_2(r)^{\frac{3}{8}}\right) \, .
\end{equation} The 6d Einstein frame metric reduces to: 
\begin{equation}\label{6dsmeared}
    \begin{aligned}
    ds^2_6 &= e^{\frac{t}{\sqrt{2}}}\left(Z_6(r)^{-\frac{1}{8}}Z_2(r)^{-\frac{5}{8}} \eta_{\mu\nu} dx ^\mu dx^\nu + Z_6(r)^{\frac{7}{8}}Z_2(r)^{\frac{3}{8}}( dr^2+r^2 d\Omega^2_2)\right)
     \\ \; &=Z(r)^{-\frac{1}{4}}\eta_{\mu\nu} dx ^\mu dx^\nu + Z(r)^{\frac{3}{4}}(dr^2+r^2 d\Omega^2_2)
    \end{aligned}
\end{equation}
where $Z(r)= Z_6(r)Z_2(r)$. 

One can see that the dilaton and radion are both blowing up upon reaching the point $r=0$, which is at finite spacetime distance, hence the configuration can be dubbed a 
6d small black 2-brane. 

As in section \ref{sec:branes}, we can describe the configuration as a Dynamical Cobordism of the 6d theory compactified on an $\IS^2$ with suitable 2-form fluxes (for the RR 2-form field strength and the $\IT^4$ reduction of the RR 6-form field strength). To implement this, we take the general ansatz: 
\begin{equation}
    ds^2_6 = e^{- 2 \alpha \sigma} ds^2_4 + r_0^2 e^{ 2 \beta \sigma} d\Omega^2_2 \, .
\end{equation}
In the resulting 4d theory, there are non-trivial potential terms for the new radion $\sigma$ arising from the curvature of $\IS^2$ and the 2-form fluxes. Imposing the Einstein frame in 4d comes down to setting $\gamma=\frac{\alpha}{\beta}=1$. One can then choose $\beta$ such that the radion $\sigma(r)$ has a canonically normalized kinetic term and one obtains $\beta=\frac{1}{2}$. From matching this compactification ansatz to equation (\ref{6dsmeared}), we obtain the canonically normalized radion $\sigma$,
\begin{equation}
    \sigma(r)=\log\left(\frac{r^2}{r_0^2 }Z(r)^{\frac{3}{4}}\right) \, ,
\end{equation} and the following 4d Einstein frame metric
\begin{equation}\begin{aligned}
    ds^2_4 &= e^{\sigma}\left(Z(r)^{-\frac{1}{4}}\eta_{\mu\nu} dx ^\mu dx^\nu + Z(r)^{\frac{3}{4}}dr^2\right)
    \; = \left(\frac{r}{r_0}\right) ^2 \left(Z(r)^{\frac{1}{2}}\eta_{\mu\nu} dx ^\mu dx^\nu + Z(r)^{\frac{3}{2}}dr^2\right)\nonumber \, .
\end{aligned}\end{equation}
This solution is a 4d Dynamical Cobordism, with the D2/D6-brane system playing the role of cobordism defect. The solution has the structure of an ETW brane; there are 3 running scalars going off to infinite distance at the singularity at $r=0$, which is straightforward to show lies at finite spacetime distance. Indeed, near $r=0$, we have
\begin{equation}
    \Delta=\int _0 ^r \left(\frac{r}{r_0}\right)(Z_6(r)Z_2(r))^{\frac{3}{4}}dr \sim \sqrt{r} \, .
\end{equation}
Furthermore, near the singularity, the distance in field space goes like: 
\begin{equation}
 dD^2 = d\Phi^2 + d\sigma^2 + d t^2 \simeq \frac{1}{2}\frac{dr^2}{r^2} \rightarrow D \simeq - \frac{1}{\sqrt{2}} \log(r)
\end{equation}
Near the singularity, the Ricci scalar in 4d behaves as:
\begin{equation}
    |R|\sim r^{-1}
\end{equation}
These lead to the familiar scaling relations near $r=0$: 
\begin{equation}\label{eq:scal-rel-D6D2}
    |R|^{-\frac{1}{2}} \sim \Delta \sim e^{-\frac{1}{\sqrt{2}}|D|} \, .
\end{equation}

\medskip

Since the above full solution has the structure of a Dynamical Cobordism, it should be possible to express it in the framework of our local description, with the D2/D6-brane system playing the role of the ETW brane. Let us define the new coordinate:
\begin{equation}
    y = \int ^r \left(\frac{r}{r_0}\right)\left(Z_6(r)Z_2(r)\right)^{\frac{3}{4}}dr \sim \sqrt{r} \, ,
\end{equation}
where we have considered the leading behaviour near $r=0$.

Using equation \eqref{radion}, we have:
\begin{equation}\label{eq:Sigma-D6/D2-T4}
    \sigma(y^2 )= -\frac{1}{2}\log\left(\frac{y^4}{r_0^2} Z(y^2)^{\frac{1}{2}}\right) \simeq - \log y \, .
\end{equation}
Matching this to the profile in \eqref{eq:sol-phi-simple}, we see that $\delta^2=2$ and  $a=\frac{2}{3}$. Then we automatically fall back on the previous field-space distance and scaling relations using equations \eqref{eq:sol-phi-simple} and \eqref{eq:scalingrelations}:
\begin{equation}
    D(y) \simeq -\sqrt{2}\log y \, ,
\end{equation}
\begin{equation}
    \Delta = y \sim e^{-\frac{1}{\sqrt{2}}\,D} \sim |R|^{-\frac{1}{2}} \, .
\end{equation}
This gives yet another nice check of the usefulness of the local analysis.

\medskip

\subsubsection*{Beyond the Einstein frame}

One last remark that will be relevant in the next sections is that scaling relations similar to those of \eqref{eq:scal-rel-D6D2} can be found, independent of the frame chosen during the compactification. Indeed, if one insists on keeping $\gamma$ (and thus, also $\beta$) general and tracking it throughout the computations, one obtains the new coordinate near $r=0$:
\begin{equation}
    \Delta = y \sim r^{\frac{1}{4}(\gamma+1)} \;\;\; \text{ and }\;\;\; |R| \sim r^{-\frac{1}{2}(\gamma+1)} \, .
\end{equation}

Note that, if $\gamma< -1$, then these scalings behave opposite to those we have seen for ETW branes. This illustrates that the scalings mentioned rely on using the Einstein frame metric to describe the ETW brane. 

In setups where one needs (or finds convenient) to use general frames, the condition for an ETW brane is that the picture of a scalar going off to infinity at finite spacetime distance can be attained by a suitable change of frame. In this respect, we note that there is an extra subtlety in dealing with the field space distance in general frames. Indeed, not being in the Einstein frame implies that the radion is multiplying the Einstein-Hilbert term in the action:
\begin{equation}\label{eq:D6D2-T4S2-action-gframe}
    S_4 \supset \frac{1}{2}\int d^4 x \sqrt{-g_4}e^{2\beta\sigma(1-2\gamma)}\{ e^{2\beta \gamma \sigma}(R_4-(\partial t)^2 -(\partial \Phi)^2  -\beta^2 (6\gamma^2-8\gamma+6) (\partial \sigma)^2) \} \, .
\end{equation} 
It thus makes sense to define the field space distance measured in units set by this coefficient of the Ricci scalar in the action. This field space distance near the singularity in this general frame reads: 
\begin{equation}\begin{gathered}
 dD^2 = d\Phi^2 +\beta^2(6\gamma^2-8\gamma+6) d\sigma^2 + d t^2 \\ D \simeq - \frac{\sqrt{6\gamma^2-8\gamma+10}}{4} \log r \, .
\end{gathered} \end{equation}
Hence, we can derive the following universal scaling relations in a general frame:
\begin{equation}
    \Delta \sim e^{-\frac{\gamma+1}{\sqrt{6\gamma^2-8\gamma+10}}|D|}\sim |R|^{-\frac{1}{2}} \, .
\end{equation}
Note that these reduce to those of \eqref{eq:scal-rel-D6D2} when setting $\gamma=1$, as required by the Einstein frame. As a side note, one cannot recover this result in the local description detailed in section \ref{sec:localdescription} as it was constructed in the Einstein frame. We leave such a more general formulation of the local construction for future work. 

\subsection{Small Black Holes from the D2/D6 system on $\IT^4\times \IT^2$}
\label{sec:D6D2-T4T2S2}

Let us now consider turning our D6/D2-brane systems into a (small) black hole, by a further compactification on $\IT^2$.  

We take the ansatz 
\begin{equation}
    ds_6^2= e^{-q}ds^2_4 + e^{q} ds^2_{T^2} \, .
\end{equation}
By matching this ansatz to the 6d metric obtained previously \eqref{6dsmeared}, we get the 4d Einstein frame metric:
\beqa
\label{eq:D6D2-T4T2-sol}
e^{q(r)}&= &Z(r)^{-\frac{2}{8}} \, , \nonumber \\
    ds^2_4&=& (g_4)_{ij} dx^i dx^j= e^{q(r)}\left(-Z(r)^{-\frac{2}{8}} dt^2 + Z(r)^{\frac{6}{8}} (dr^2+r^2 d\Omega^2_2)\right)\nonumber \\
 &\;= &-Z(r)^{-\frac{1}{2}} dt^2 + Z(r)^{\frac{1}{2}} (dr^2+r^2 d\Omega^2_2)\; .
\eeqa 
This solution describes a small black hole (in fact, equivalent to the celebrated D1/D5-brane one, by T-duality in one of the $\IT^2$ directions), of the kind considered in \cite{Hamada:2021yxy}. 

To motivate the relation with the more general discussion in the next section, let us make the following heuristic argument. 
Although our solution has three scalar fields, the radial evolution can be reduced to one effective scalar as follows.
Near $r=0$, all three scalars have the same profile, so we may combine them in one effective scalar $D$ whose effective action near $r=0$ is of the form
\begin{equation}\label{eq:D6/D2-T4S2-action-simple}
    S_4 \sim \frac{1}{2}\int d^4 x \sqrt{g_4}\{R_4 - (\partial D) ^2- \frac{1}{4}e^{\frac{1}{ \sqrt{2}}D}|F_2|^2\}
\end{equation}
where we have restricted to the $U(1)$ linear combination under which the D2/D6 system is charged. 

With this proviso, we can frame this particular example with the more general class of small Black Holes considered in \cite{Hamada:2021yxy}, to be discussed next.

\subsection{General small Black Holes}
\label{sec:hmvv-smallBHs}

In the context of the swampland program, \cite{Hamada:2021yxy} proposed the use of 4d small black hole solutions to provide further evidence for a number of a number of Swampland conjectures. A particularly important property is that the 4d solutions contain scalars going off to infinite field space distance at the black hole core. In the spirit of previous sections, in this section we show that these 4d solutions can be turned into 2d Dynamical Cobordisms upon reducing on the $\IS^2$, with the small black hole playing the role of the ETW brane. In fact we will check that the 2d running solution satisfies the familiar scaling relations (for a general frame, since there is no Einstein frame in 2d).

Let us briefly review the key features of such solutions.
We consider 4d Einstein-Maxwell coupled to a scalar controlling the gauge coupling. We take the action
\begin{equation} \label{eq:4d-action}
	S_{4d} \sim \frac{1}{2}\int d^{4}x\, \sqrt{-g_4}\,  \left( R_{4} -  \left( \partial \phi\right)^{2} - e^{2a\phi} |F_2|^{2} \right) \, . 
\end{equation}
We focus on exponential dependence, since it provided the most explicit class considered in \cite{Hamada:2021yxy}. It also fits with the special role of exponential functions in local descriptions of ETW branes.

Without loss of generality, we take $a>0$ so that $\phi\to\infty$ corresponds to weak coupling for the $U(1)$ gauge field. Note that this $a$ should not be confused with the parameter in \eqref{eq:enter-a}, and we trust the reader to distinguish them by the context.

In this theory, electrically charged extremal black holes take the form
\begin{equation} \label{eq:4d-BH-sol}
	ds_{4}^{2} = - f(r) dt^{2} + f(r)^{-1} dr^2 + r^2 R(r)^2 d\Omega_{2}^{2} \, ,
\end{equation}
where
\begin{equation}
    R(r) = \left( 1 - \frac{r_{h}}{r} \right)^{\frac{a^2}{1+a^2}} \, , \quad \,	f(r) = \left( 1 - \frac{r_{h}}{r} \right)^{\frac{2}{1+a^2}} \, .
\end{equation}
In addition, the profile for the scalar is given by
\begin{equation} \label{eq:profile-phi}
	\phi(r) = \phi_{0} - \frac{\sqrt{2} \,a}{1+a^2} \log \left( 1 - \frac{r_h}{r}\right) \, .
\end{equation}

The scalar goes off to infinity at the horizon $r=r_h$, which is however not smooth, since the $\IS^2$ shrinks to zero size, leading to a small black hole.

In the string theory context, small black holes can be easily built by using D-branes. In fact, we now recast the above solution in a form closer to the solution \eqref{eq:D6D2-T4T2-sol}, which described our system of D2- and D6-branes on $\IT^4\times \IT^2$. This was already anticipated when we obtained \eqref{eq:D6/D2-T4S2-action-simple}, which has the structure of \eqref{eq:4d-action} (for $a=\frac{1}{2\sqrt{2}}$).

Carrying out the coordinate change $r \rightarrow r+r_h$, the metric \eqref{eq:4d-BH-sol} becomes
\begin{equation}\label{eq:4d-BH-sol-nice}
    ds^2_4 = -\left(1+ \frac{r_h}{r}\right)^{-\frac{2}{1+a^2}}dt^2 + \left(1+ \frac{r_h}{r}\right)^{\frac{2}{1+a^2}}\left(d r^2 + r ^2 d\Omega_2^2\right) \, .
\end{equation}
Similarly, the scalar reads 
\begin{equation}\label{eq:4d-BH-field-nice}
    \phi(r)= \phi_0 + \frac{\sqrt{2} \,a}{1+a^2} \log \left(1+\frac{r_h}{r}\right) \, .
\end{equation}
This has the structure of \eqref{eq:D6D2-T4T2-sol} with $Z(r)=(1+r_h/r)^{\frac{4}{1+a^2}}$. Note that the core of the small black hole now lies at $r=0$.

We now perform the reduction on $\IS^2$ to express these solutions as 2d running solutions describing a local Dynamical Cobordism, with the small black hole playing the role of the ETW brane. We will also recover the corresponding (general frame) scaling relations.

Since there is no Einstein frame in 2d, we perform the $\IS^{2}$ reduction with the following general ansatz:
\begin{equation} 
\label{eq:compact2d-ansatz}
	ds_{4}^{2} = e^{-2\alpha\omega} ds_{2}^{2} + e^{2\beta\omega} r_{0}^{2} d\Omega_{2} \, .  
\end{equation}
The 2d action obtained from the compactification contains the terms
\begin{equation} \label{eq:2d-action}
	S_{2d} \supset \frac{1}{2} \int d^{2}x\, \sqrt{-g_2}\, e^{2\beta\omega} \left( R_{2} - \left( \partial \phi\right)^{2} - 6 \beta^2 \left(\partial\omega\right)^{2} \right) \, .
\end{equation}
These expressions already show the impossibility to define an Einstein frame: it would require $\beta=0$, and this would kill the radion's kinetic term. We therefore keep $\beta$ general, so we deal with a dilaton-gravity theory. By matching the ansatz \eqref{eq:compact2d-ansatz} with the 4d metric \eqref{eq:4d-BH-sol-nice} we get the profile for the radion
\begin{equation} \label{eq:profile-radion}
	\omega(r) = \frac{1}{\beta} \log \left( \frac{r}{r_0}\left(1+ \frac{r_h}{r}\right)^{\frac{1}{1+a^2}}\right) \, ,
\end{equation}
and the 2d metric
\begin{equation}
	ds_{2}^{2} = \left( \frac{r}{r_0}\right)^{2\gamma} \left( -\left(1+ \frac{r_h}{r}\right)^{-\frac{2(1-\gamma)}{1+a^2}}dt^{2} +\left(1+ \frac{r_h}{r}\right)^{\frac{2(1+\gamma)}{1+a^2}}dr^2\right) \, ,
\end{equation}
where $\gamma=\frac{\alpha}{\beta}$.

Computing the 2d Ricci scalar and taking the leading order in $r\to 0$ we get
\begin{equation}
	|R| \sim r^{-2\frac{(\gamma+1)a^2}{1+a^2}} \, ,
\end{equation}
where we are ignoring a constant prefactor\footnote{This prefactor vanishes for either $a^2=1$ or $a^2=-2\gamma$. We will skip these cases without further discussion.}. 

Similarly, the spacetime distance from a given $r$ to the singularity, at leading order in $r\to 0$, scales as
\begin{equation}
	\Delta \sim r^{\frac{(\gamma+1)a^2}{(1+a^2)}} \, .
\end{equation}
We note that, as expected, the scaling is the familiar ETW one if $\gamma> -1$. As explained above, the fact that 2d gravity is topological means that the criterion for an ETW brane in a solution should be that the usual relations hold in {\em some} suitable frame.

Let us now recover the usual scalings with the field distance. Recalling the latter is measured in units set by the coefficient of the Ricci scalar in the action, we can it read off from \eqref{eq:2d-action} as: 
\begin{equation}
	dD^{2} = d\phi^{2} + 6 \beta^2 d\omega^{2} \, .
\end{equation}
Plugging the profiles \eqref{eq:profile-phi} and \eqref{eq:profile-radion} at leading order as $r\to 0$ and integrating the line element we recover
\begin{equation}
	D(r) \simeq - \frac{a \sqrt{2+6a^2}}{1+a^2}\, \log r \,.
\end{equation}
Finally, together with the previous results for the distance to the end of the world and the curvature, we obtain the scalings
\begin{equation}
    \Delta \sim e^{-\frac{\delta}{2} \,D} \, , \quad |R| \sim e^{\delta \,D} \, ,
\end{equation}
with
\begin{equation}
	\delta = \frac{2(\gamma + 1) a}{\sqrt{2+6a^2}} \, .
\end{equation}
Hence, we recover the general frame scaling relations introduced in section \ref{sec:D6D2-T4S2}. This shows that small black hole solutions can be regarded as just another instance of Dynamical Cobordism, and that they admit local scaling relations identifying the small black hole core with ETW branes in 2d. 

\section{Swampland constraints and Surprises from the UV}
\label{sec:surprise}

In this section we discuss interesting interplays of the scalar running off to infinity in field space in Local Dynamical Cobordisms and the Swampland constraints.

\subsection{Swampland Distance Conjecture and other constraints}
\label{sec:swampland}

Many studies of Swampland constraints are related to infinity in scalar moduli/field space (see \cite{Brennan:2017rbf,Palti:2019pca,vanBeest:2021lhn} for reviews). Since Dynamical Cobordisms explore infinite field space distances, in this section we discuss the interplay with different Swampland constraints, especially the Distance Conjecture \cite{Ooguri:2006in} (see \cite{Nicolis:2008wh,Baume:2016psm,Valenzuela:2016yny,Heidenreich:2017sim,Ooguri:2018wrx,Heidenreich:2018kpg,Grimm:2018ohb,Corvilain:2018lgw,Buratti:2018xjt,Lust:2019zwm,Grimm:2019ixq,Gendler:2020dfp,Calderon-Infante:2020dhm,Baume:2020dqd,Perlmutter:2020buo} and the reviews above for other approaches).

Let us focus on the simplest expression of the Distance Conjecture, which states that, when the scalars are taken to infinite field space distance $D$ (in an adiabatic approach, namely, by changing the spacetime independent vevs), there is a tower of states becoming exponentially light, and thus the cutoff of the effective theory is lowered as
\beqa
\Lambda\sim e^{-\alpha D} \, ,
\eeqa 
with some positive order 1 coefficient $\alpha$. 

This scaling can be combined in an interesting way with our scalings near ETW branes. For instance, using \eqref{eq:scalingrelations}, we have
\beqa
\Lambda\sim \Delta^{\frac {2\alpha}\delta} \, .
\eeqa 
This matches with our intuition that the full description of the ETW brane requires UV completing the effective theory. It is important to note that  the appearance of an infinite tower in the adiabatic version of the Distance Conjecture does not necessarily imply the appearance of a tower in the present Dynamical Cobordism context. On the other hand, the lowered cutoff certainly signals that there could be situations where the naive ETW brane picture as described in effective theory may be corrected. We will see explicit examples in section \ref{sec:largeN}.

Using also \eqref{eq:scalingrelations}, we get that the cutoff scale relates to the spacetime curvature as
\beqa 
|R|\sim \Lambda^{-\frac{\delta}{\alpha}} \, ,
\eeqa 
(where we have taken the generic case $\delta\neq (2d/(d-2))^{1/2}$ for concreteness). This relation, already noted in \cite{Buratti:2021fiv} is reminiscent of (although admittedly different in spirit from) that in \cite{Lust:2019zwm} for AdS vacua.

From this perspective, the correlation between the appearance of the naked singularity and the running of the scalar going off to infinity suggests that the lowered cutoff of the swampland distance conjecture is responsible for regulating the singularity, which would be resolved in a more complete microscopic UV description. This remark is in the spirit of \cite{Lanza:2021qsu} (see also \cite{Lanza:2020qmt}) and \cite{Hamada:2021yxy}, where the singular behaviour of certain defects (EFT strings or small black holes, respectively) is related to scalars going off to infinite distance. 

From our perspective, the relation follows from the Dynamical Cobordism Distance Conjecture in \cite{Buratti:2021fiv}. In our present terms: Every infinite field distance limit of an effective theory consistent with quantum gravity can be realized as a solution running into a cobordism ETW brane (possibly in a suitable compactification of the theory).

In particular, in Sections \ref{sec:branes} and \ref{sec:smallBHS} we 
provided a description of general defects as ETW branes of Dynamical Cobordisms. This general framework encompasses the defects in \cite{Lanza:2021qsu,Hamada:2021yxy} as particular examples.

\medskip

An interesting spin-off of our local analysis is that it constrains the asymptotic form of the potential. Namely, whenever it is not vanishing (actually, negligible as compared with the scalar kinetic energy) it has an exponential form with a critical exponent $\delta$, c.f. \eqref{eq:sol-V}. It is thus interesting to compare this asymptotic form of the potential with Swampland constraints expected to hold near infinity in scalar field space.

Let us consider the de Sitter conjecture in the version of \cite{Obied:2018sgi} (see \cite{Garg:2018reu,Ooguri:2018wrx} for the refined one), namely $|\nabla V|/V > {\cal O}(1) $. From \eqref{eq:sol-V} we have
\beqa
\frac {V'}V=\delta \, .
\eeqa 
Since in general the critical exponent $\delta\sim {\cal O}(1)$, the potential satisfies the de Sitter conjecture. This fits nicely with the idea that the latter is expected to hold near infinity in moduli/field space.

Moreover, let us compare with the Transplanckian Censorship Conjecture \cite{Bedroya:2019snp}
\beqa
{|\nabla V|}\geq \frac{2}{\sqrt{(d-1)(d-2)}}\, V \, .
\eeqa
When $V<0$, the constraint is trivial; on the other hand, when $V>0$, in our setup we must have $a< 0$, and the expression \eqref{eq:gamma} for $\delta$ guarantees that the above inequality is satisfied.
A caveat for the above statements is that both the de Sitter and the Transplanckian Censorship conjectures involve the gradient $\nabla V$, whereas our local description provides the potential only along one direction, the effective scalar dominating the running near the ETW brane. Hence, the comments above would hold under the assumption that the effective scalar in the local description follows a gradient flow. It would be interesting to assess this point in explicit models, and we leave this as an open question for future work.

\subsection{Large $N$ surprises from the UV}
\label{sec:largeN}

In the previous section we have discussed that the Distance Conjecture implies a lowered cutoff as one approaches the ETW brane. Indeed, as mentioned at several points, the microscopic description of the ETW branes lies in the underlying UV completion. In most of our examples, the corresponding cobordism defect is known, so that the end of the world picture can be confirmed in the full theory. However, it is conceivable that in some specific cases there exist UV effects hidden at the core of the ETW brane potentially modifying this picture. In this section we present two examples, where such corrections exist and lead to large backreactions, ultimately turning the candidate ETW brane into a domain wall interpolating to a new region beyond the apparent singularity. A further interesting observation is that both examples are related to large $N$ physics and holography.

\medskip

{\bf Large number of M2-branes}

Consider as our first example a stack of $N$ D2-branes in flat 10d spacetime (or at a smooth point in any other compactification). Locally around the D2-brane location the $\IS^6$ truncation yields a 4d theory with an ETW brane, at which a scalar (a combination of the radion and the dilaton) goes to infinity in field space. One may follow the theory in this limit and, as noted in \cite{Itzhaki:1998dd}, realize that the strong coupling is solved by lifting to M-theory, and turning the D2-branes into M2-branes. For small $N$, the UV completion of the effective ETW brane is thus merely a stack of M2-branes removing the flux and allowing spacetime to end, as befits a Dynamical Cobordism. 

On the other hand, for $N$ large we have a different behavior: the large number of M2-branes backreact on the geometry and generate an infinite AdS$_4\times \IS^7$ throat. The effective theory ETW brane has a UV description with so many degrees of freedom that it actually generates a gravity dual beyond the wall. 

From the perspective of the running scalars, the AdS$_4\times\IS^7$ represents a minimum of the ($\IS^7$ radion) potential. Hence the full D2/M2 solution describes the running  of the theory from the slope of the potential down to a stable minimum, at which the theory relaxes to a maximally symmetric solution, instead of hitting an end of the world. The location of the minimum in field space is hidden near infinity in the original D2-brane effective description. Hence, the large $N$ allows for the appearance of a minimum at strong coupling, which is nevertheless tractable\footnote{This is reminiscent of the argument \cite{Buratti:2020kda} that the scale separation (and hence the tractability) of the AdS minima in \cite{DeWolfe:2005uu,Camara:2005dc} is controlled by a large number of flux units.}. 

Moreover, the full D2/M2 solution describes a dynamical cobordism from the M-theory perspective. Far away from the stack of branes we can use the description in terms of D2-branes. As described above the 4d theory would be obtained by compactifying Type IIA on an $\IS^6$. This would be further lifted to M-theory on $\IS^6\times\IS^1$. On the other hand, we have just argued that close to the stack of branes the 4d theory is given by M-theory on $\IS^7$. We then see that this solution describes a dynamical cobordism between to different compactifications. Notice that this is not a cobordism to nothing, described by ETW brane solutions.

\medskip

{\bf Warped KS throat with large number of D3-branes}

Our second example is based on the warped throat considered in section \ref{sec:ks}. Recall we have type IIB theory compactified on $T^{1,1}$ with $N$ units of RR 5-form flux and $M$ units of RR 3-form flux on the $\IS^3$, and we focus on the choice of parameters $N=KM+P$. At the level of the 4d effective theory, we recover a KT solution with a singularity at a finite spacetime distance, at which a scalar (a combination of the $T^{1,1}$ radion and the dilaton, but dominated by the former) goes off to infinite field space distance.

The UV smoothing of this singularity is slightly trickier than the $N=KM$ case of section \ref{sec:ks}. It involves the smoothing of the singular conifold geometry into a deformed conifold, with a finite size $\IS^3$, but there remain $P$ D3-branes at the tip of the throat. This can be shown using the holographic dual field theory, as follows. There is a Seiberg duality cascade from the initial $SU(N)\times SU(N+M)$ theory in which $N$ effectively decreases in multiples of $M$; hence, in the last step of the cascade we have an $SU(P)\times SU(M+P)$ gauge theory, whose strong coupling dynamics leads to an remnant $\NN=4$ $SU(P)$ theory, as befits the above mentioned $P$ probe D3-branes.

Hence, for small $P$ the ETW brane of the 5d theory is microscopically described by the smooth Klebanov-Strassler throat dressed with $P$ explicit D3-branes, required to absorb the remnant 5-form flux and allow spacetime to end.

On the other hand, for $P$ large we have a different behavior: the large number of D3-branes backreact on the geometry and generate an infinite AdS$_5\times \IS^5$ throat. The effective theory ETW brane has a UV description with so many degrees of freedom that it actually generates a gravity dual beyond the wall. The interpretation of this strong correction in terms of the running scalars is similar to the one mentioned above, as the apperance of an AdS minimum hidden near the infinite field space distance limit of the effective description.

\medskip

We have seen two examples in which a naive ETW brane in the effective description has a UV description encoding large backreactions on the geometry recreating a geometry beyond the wall. Alternatively, the corrections generate minima in the scalar potential in the region near field space infinity of the effective description. It would be interesting to explore in more detail these and other possible classes of examples exhibiting this phenomenon. We hope to report on this in the future.

\medskip

\section{Conclusions}
\label{sec:conclusions}

In this paper we have studied Dynamical Cobordism solutions in which theories of gravity coupled to scalars develop an end of spacetime. The latter is encoded in the effective theory as the appearance of a singularity at finite spacetime distance, at which some scalars run off to infinite field space distance. We have provided a local description of the configurations in the near ETW brane regime, and shown that the solutions are largely simplified, and fall in universality classes characterized by a critical exponent $\delta$, which controls the profiles of the different fields and the scaling relations among the field space distance $D$, spacetime distance $\Delta$ and scalar curvature $R$.
 
We have studied several explicit models of ETW branes and characterized them in the local description, computing their critical exponent. The different examples and their key parameters are displayed in Table \ref{tab:table-examples}. This list is intended to illustrate typical values of these parameters. It would be interesting to explore more examples and to explore possible connections among ETW branes described by the same parameters.

\medskip

\begin{table}[htb] \footnotesize
{\renewcommand\arraystretch{1.4}
\begin{center}
\begin{tabular}
{|c|c|c|c|}
\hline
\textbf{Example} & $d$ & $\delta$ & $a$  \\
\hline
\textit{Massive IIA} & $10$ & $\frac{5}{\sqrt{2}}$ & $- \frac{16}{5}$ \\
\hline
\textit{Non-susy $USp(32)$ string} & $10$ & $\frac{3}{ \sqrt{2}}$ & $0$ \\
\hline
\textit{D7 branes} & $9$ & $\frac{4 \sqrt{14}}{7}$ & $0$ \\
\hline
\textit{D6 branes} & $8$ & $\sqrt{2}$ & $\frac{4}{7}$ \\
\hline
\textit{D5 branes} & $7$ & $\frac{2 }{\sqrt{5}}$ & $\frac{5}{6}$ \\
\hline
\textit{D4 branes} & $6$ & $\frac{1}{\sqrt{5}}$ & $\frac{24}{25}$ \\
\hline
\textit{Klebanov-Strassler} & $5$ & $\frac{2 \sqrt{30}}{3}$ & $- \frac{3}{2}$ \\
\hline
\textit{Bubble of Nothing} & $4$ & $\sqrt{6}$ & $0$ \\
\hline
\textit{D2 branes} & $4$ & $\frac{\sqrt{14}}{7}$ & $\frac{20}{21}$ \\
\hline
\textit{D2/D6 on $T^4 \times	S^2$} & $4$ & $\sqrt{2}$ & $\frac{2}{3}$\\
\hline
\textit{D1 branes} & $3$ & $\sqrt{2}$ & $\frac{3}{4}$ \\
\hline
\textit{EFT string} & $3$ & $2 \sqrt{2}$ & $0$ \\
\hline
\end{tabular}
\end{center}
}
\caption{Table of examples in this paper, with the corresponding parameters for the local description near the ETW brane.}
\label{tab:table-examples}
\end{table}


We have moreover shown that small black holes can also be regarded as Dynamical Cobordisms, and satisfy similar scaling laws. It would be interesting to explore from the cobordism perspective the recent applications of small black holes to the derivation of swampland constraints.

There are several interesting open directions for the future:

$\bullet$ We have focused on solutions with spatial dependence. It would certainly be interesting to explore time-dependent backgrounds, and their possible application to cosmology.

$\bullet$ In our local analysis we have focused on certain particular choices. For instance, we have not considered solutions where $|V|\gg |V_t|$, and we have moreover taken solutions controlled by a constant parameter $a<1$. More general possibilities are in principle allowed from a mere effective field theory perspective, but they are not realized in any of the string theory examples we have explored. It is thus an interesting question if there are UV complete models realizing them, or on the contrary, they are excluded by some further arguments of consistency with Quantum Gravity.

$\bullet$ Finally, it would be interesting to get a better understanding of the possible appearance of non-trivial corrections in the large field region near the ETW branes, in particular those leading to large backreactions signalling the existence of new minima of the scalar potential. This could lead to further insights into the stabilization of moduli in asymptotic regions of moduli/field space. The two examples mentioned in our work signal an interesting interplay with large $N$ limits and holography, which may provide an extra leverage on these configurations.

We hope our work motivates interesting results in this and other directions.

%
\section*{Acknowledgments}
We are pleased to thank I. Basile, J. R. Espinosa, L. Ib\'anez, F. Marchesano, M. Montero and I.Valenzuela for useful discussions. J.C. and M.D. also wish to acknowledge the hospitality of the Department of Physics of Harvard University during the early and late stages of this work, respectively. This work is supported through the grants CEX2020-001007-S and PGC2018-095976-B-C21, funded by MCIN/AEI/10.13039/501100011033 and by ERDF A way of making Europe. The work by R.A. is supported by the grant BESST-VACUA of CSIC.  The work by M.D. is supported by the FPI gran no. FPI SEV-2016-0597-19-3 from Spanish National Research Agency from the Ministry of Science and Innovation. The work by J.C. and J. H. is supported by the FPU grants no. FPU17/04181 and FPU20/01495 from the Spanish Ministry of Education.

\newpage

\appendix

\section{Local Dynamical Cobordisms with curved $(d-1)$-dimensional slices}
\label{app:curvature}

\subsection{General analysis for curved slices}
\label{sec:general-curved}

We can generalize the discussion in Section \ref{sec:local} to the case in which the ETW brane has constant internal curvature $R_d$. Namely we take the foliation ansatz \eqref{dw-ansatz} with $ds_{d-1}^{2}$ describing a constant curvature $(d-1)$-dimensional metric. The equations of motion read
\begin{align} 
	&(d-1) \sqrt{2\left(V-V_{t}\right)} \, \sigma^{\prime} - \partial_{\phi}V_{t} = 0 \, , \label{eq:app-eoms}\\
    &\frac{1}{2}(d-1)(d-2) \sigma^{\prime\,2} + V_{t} -\frac{1}{2}e^{2\sigma}R_d= 0 \, ,  \\
    &(d-2) \sigma^{\prime\prime} - 2\left(V-V_{t}\right) -\frac{1}{d-1}e^{2\sigma}R_d= 0 \label{eq:app-check}\, ,
\end{align}
where we have again introduced the tunneling potential defined in \eqref{eq:tunneling-potential}.

For $R_d\neq0$, it is still possible to eliminate $\sigma$ by combining the first two equations (and their derivatives): 
\begin{equation}\label{app:eom}
	\left(d\,\partial_{\phi}V_t-(d-1)\partial_{\phi}V\right) \partial_{\phi}V_t=2(d-1)(V_t-V) \left[ \partial_{\phi}^{2}V_t+\frac{2}{d-2} ((d-1)V-(d-2)V_t)\right] \, .
\end{equation}
Importantly, in this derivation we need to assume $R_d\neq 0$, so that we do not expect to necessarily recover the results in section \ref{sec:local}.

Restricting to the case $V=aV_t$, with $a$ a constant, we find that the solution to this equation is
\begin{equation} \label{eq:app-sol-Vt}
    V_t(\phi)= - c \left( \cosh \left(\frac{  (a (d-1)+2-d)\phi}{\sqrt{(1-a) (d-2) (d-1)}}\right)\right) ^{2-\frac{2}{a (d-1)+2-d}} \, ,
\end{equation}
where we have ignored an integration constant that is irrelevant for the $\phi\to\infty$ limit. 

Notice that, for $a>1$, the coefficient in front of $\phi$ becomes imaginary and then what we have is a cosine, rather than a hyperbolic cosine. As we are not interested in this behaviour we from now on require $a<1$. From computing $\phi'^2$ from this solution and requiring that it must be positive, we then learn that we must have $c>0$.

In addition, as we are interested in ETW branes, we want to require that $\phi'^2$ blows up as $\phi\to\infty$. This is equivalent to having $|V_t|\to\infty$ in this same limit, which in turn implies that the power in \eqref{eq:app-sol-Vt} must be positive. This gives us that the only ETW brane solutions are for $a<\frac{d-2}{d-1}$. For this range of $a$, we can approximate the hyperbolic cosine by an exponential (as we are interested in the limit $\phi\to\infty$) and we have
\begin{equation}
    V_t(\phi) \simeq - c \left( \exp \left(\frac{  (a (d-1)+2-d)\phi}{\sqrt{(1-a) (d-2) (d-1)}}\right)\right) ^{2-\frac{2}{a (d-1)+2-d}} = - c\, e^{\delta\,\phi}\ .
\end{equation}
The coefficient $\delta$ is
\beqa 
\delta=2\sqrt{\frac{d-1}{d-2}(1-a)}\,.
\label{eq:gamma-bis}
\eeqa 
So for $a<\frac{d-2}{d-1}$ the case of a ETW brane with internal curvature coincides with the case studied in the paper. Interestingly, this case turns out to be more restrictive than the $R_d=0$ one, for which any $a<1$ described an ETW brane.

This solution was also assuming that $a\neq\frac{d-2}{d-1}$. Plugging that particular value in \eqref{app:eom}, we find that the equation of motion simplifies to
\begin{equation}\label{app:eom2}
	(\partial_\phi V_t)^2=V_t\cdot \partial_\phi^2 V_t \, .
\end{equation}
This equation has the solution
\begin{equation}
	V_t = - c\, e^{\delta \, \phi} \, ,
\end{equation}
with $c$ and $\delta$ arbitrary constants. In order to describe an ETW brane we require $\delta>0$. Interestingly, for this special value of $a$ with $R_d\neq0$, we find that we recover the exponential behaviour, but with the freedom of choosing the critical exponent $\delta$.

In both cases we find the same exponential behaviour for $V_t$. Therefore, just as in section \ref{sec:localdescription}, we find that the potential takes the form
\begin{equation}
	V(\phi) \simeq - a\,c \, e^{\delta\, \phi} \, .
\end{equation}
However, here we uncover that, for a given potential of this form, the setup with $R_d\neq 0$ allows for two possible values of $a$, namely the $a<\frac{d-2}{d-1}$ given in \eqref{eq:gamma-bis}), or the value $a=\frac{d-2}{d-1}$, with $\delta$ and $a$ independent. For this reason, from now on we keep $a$ and $\delta$ as different variables when solving the rest of the equations, and at the end we comment on the two possibilities.

Using \eqref{eq:tunneling-potential} we can obtain the profile for $\phi$
\begin{equation}
	\phi(y) \simeq - \frac{2}{\delta} \log \left( \frac{\delta}{2} \sqrt{2(1-a)c} \, y \right) \, .
\end{equation}
Notice that this is the equivalent to \eqref{eq:sol-phi-full}, but with $a$ and $\delta$ kept independent. The leading behaviour is then given by
\begin{equation}\label{field:appendix}
    \phi(y) \simeq -\frac{2}{\delta}\log y \, ,
\end{equation}
and thus the field only depends on the critical exponent.

We can now use \eqref{eq:app-eoms} to get the profile for the warp factor $\sigma$:
\begin{equation}
    \sigma \simeq - \frac{1}{(d-1)(1-a)} \log y \, ,
    \label{eq:sigma-curved}
\end{equation}
where we have set an integration constant to zero without loss of generality. We recover the equivalent to \eqref{eq:sol-sigma}, albeit  with $a$ and $\delta$ kept independent. We see that the warp factor doesn't depend on $\delta$, but specifically on the prefactor $a$ of the potential.

Finally, we have to check that  the solution is compatible with \eqref{eq:app-check}. From it we obtain the condition
\begin{equation}
    \frac{4}{\delta^2} - \frac{d-2}{(d-1)(1-a)} + \frac{R_d}{d-1}\, y^{2-\frac{2}{(d-1)(1-a)}} = 0 \, .
\end{equation}
Let us now apply it for the two possible values for $a$: \\
$\bullet$ For $a<\frac{d-2}{d-1}$, the power of $y$ in the last term is positive, so that it is subleading in the $y\to 0$ limit. Moreover, recall that in this case $\delta$ relates to $a$ via \eqref{eq:gamma-bis}, which is the precise the value for which the first two terms cancel each other. In conclusion, for $a<\frac{d-2}{d-1}$ having $R_d\neq 0$ becomes irrelevant as we approach the ETW and we basically recover the same results as in the $R_d=0$ case. \\
$\bullet$ For $a=\frac{d-2}{d-1}$,  the exponent of $y$ vanishes , and hence the $R_d$ term is relevant. In this case, consistency of the equations requires
\begin{equation} \label{eq:app-delta-2}
    \delta = 2 \left( d-2 - \frac{R_d}{d-1} \right)^{-\frac{1}{2}} \, .
\end{equation}
Therefore, for this case $\delta$ is also fixed, but in terms of $R_d$. Notice that this quantity must satisfy $R_d<(d-2)(d-1)$. Provided this condition, we find that $\delta$ can take any positive value.

This case corresponds to a metric $ds^2=dy^2+y^2ds_{d-1}^2$, hence it describes a conical singularity. The singularity is absent in the case $R_d=(d-1)(d-2)$, namely the curvature of $ds_{d-1}^2$ is that of $\IS^{d-1}$, and the geometry is locally smooth, and we have $\delta=0$ and no exponential growth of the potential. Also, in order to have an ETW brane, the $(d-1)$-dimensional curvature must be lower than that of $\IS^{d-1}$.

\medskip

In conclusion, given a potential with an exponential behaviour as $\phi\to\infty$, in the $R_d\neq0$ case there exist two different kind of solutions. In the first one the value of $R_d$ is irrelevant and we recover the same behaviour as in the $R_d=0$ case (but with a more constrained critical exponent, $\delta>\frac{2}{\sqrt{d-2}}$). In the second, the curvature $R_d$ is relevant and it must be fixed by the critical exponent by \eqref{eq:app-delta-2}.

\subsection{Witten's Bubble of Nothing}
\label{sec:bon}

To illustrate the above general formulation for curved $(d-1)$-dimensional slices, we consider the example of the celebrated Witten's bubble of nothing \cite{Witten:1981gj} (see \cite{Ooguri:2017njy,GarciaEtxebarria:2020xsr,Dibitetto:2020csn,Bomans:2021ara} for other recent realization of bubbles of nothing). We show it admits a description in an effective 4d theory of gravity coupled to a scalar with zero potential, as a 4d Dynamical Cobordism, and characterize its local description and critical exponent $\delta$. 

Related discussion of a 4d effective description of the configuration have appeared in \cite{Dine:2004} (recently revisited in the context of bubbles in de Sitter space in \cite{Draper:2021ujg,Draper:2021qtc}).

Since we have restricted our discussion to dependence on spatial coordinates, we actually consider the euclidean 5d Schwarzschild black hole solution, before the Wick rotation to the expanding bubble solution. The 5d metric reads
\begin{equation}
    ds^2=\left(1-\frac{R^2}{r^2}\right)^{-1}dr^2+r^2d\Omega_3^2+\left(1-\frac{R^2}{r^2}\right)d\phi^2 \, .
\end{equation}
Here $\phi$ parametrizes an $\IS^1$ fibered over the radial coordinate $r$, times and $\IS^3$; the radial coordinate is constrained to the range $r\geq R$, and the $\IS^1$ shrinks to zero size at the euclidean horizon $r=R$ (in a smooth way for the periodicity $\phi\sim \phi + 2\pi R$).

We would like to perform a reduction to 4d along the $\IS^1$. This is a sphere reduction analogous to those in Section \ref{sec:compactification}. Hence, we match this metric with  \eqref{eq:ansatz}, for $n=1$, $d=4$, and, using \eqref{EinsteinFrame}, $\alpha=-\sqrt{1/6}$ and $\beta=-\sqrt{2/3}$. We obtain that the radion $\omega$ in \eqref{eq:ansatz} is:
\begin{equation}
    \omega=-\sqrt{\frac{3}{8}} \log \left(1-\frac{R^2}{r^2}\right) \, .
\end{equation}
The 4d metric is given in \eqref{compactansatz} and reads
\begin{equation}
    ds_4^2=\left(1-\frac{R^2}{r^2}\right)^{-\frac{1}{2}}dr^2+\left(1-\frac{R^2}{r^2}\right)^{\frac{1}{2}}r^2d\Omega_3^2 \, .
\end{equation}
We would now like to zoom into the location of the ETW brane, the euclidean horizon $r=R$. So we introduce the coordinate $\tilde{r}=1-\frac{R^2}{r^2}$. Near $r\to R$ the metric scales as
\begin{equation}\label{metric-appendix}
    ds_4^2\sim \tilde{r}^{-\frac{1}{2}}d\tilde{r}^2+\tilde{r}^{\frac{1}{2}}d\Omega_3^2 \, .
\end{equation}
Now, we make the change \eqref{y-transformation}:
\begin{equation}
    y=\int \frac{d\tilde{r}}{\tilde{r}^{1/4}}\simeq \tilde{r}^{3/4} \, .
\end{equation}
Replacing $\tilde{r}\simeq y^{\frac{4}{3}}$ in \eqref{metric-appendix} we get the 4d metric as a foliation of $\IS^3$ slices:
\begin{equation}
    ds_4^2\sim dy^2+y^{\frac{2}{3}}d\Omega_3^2 \, .
\end{equation}
This corresponds to a metric of the kind \eqref{dw-ansatz} for curved 3d slices, namely of the kind studied in appendix \ref{sec:general-curved}.
Using \eqref{eq:sigma-curved} we can see that $a=0$, and from \eqref{eq:gamma-bis}  $\delta=\sqrt{6}$. Interestingly, this corresponds to the case in which the curvature of the slices is irrelevant, and the solution is similar to the $R_d=0$ case.

We could have also obtained the same result from the profile for the radion, 
\begin{equation}
    \omega=-\sqrt{\frac{3}{8}} \log \tilde{r}\simeq -\sqrt{\frac{2}{3}} \log y\; .
\end{equation}
By using \eqref{field:appendix}, $\omega\simeq -\frac{2}{\delta}\log y$, we read that $\delta=\sqrt{6}$, hence $a=0$.

Hence Witten's bubble of nothing is described by a 4d Dynamical Cobordism running solution with the scalar reaching off to infinite distance in fields space at a rate controlled by the critical exponent $\delta=\sqrt{6}$. This provides a simple local description in terms of an ETW brane. From this perspective, the 5d solution provides the UV completion of the ETW brane, which in this case is purely a geometrical closing-off of the geometry. 

We would like to emphasize that this example provides an explicit realization of the picture discussed in Section \ref{sec:branes}, in particular Figure \ref{fig:brane} (albeit, with no brane dressing at the tip). Namely, the complete solution involves a genuine compactification on a finite size $\IS^1$, yet it is described by a local EWT brane model identical to that obtained as an $\IS^1$ reduction on a flat $\IR^2$ (which, given the vanishing potential, straightforwardly leads to $a=0$, hence $\delta=\sqrt{6}$). This supports the picture in Section \ref{sec:branes} that the sphere reductions in the flat space transverse to the D-branes suffices to provide the local description even in the (physically more interesting case) in which the transverse space is globally given by a more involved geometry, implementing the actual compactification to the lower-dimensional theory.

\section{Subleading corrections to the local description}
\label{app:subleading}

In section \ref{sec:localdescription} we took constant $a$ as a proxy for the leading behaviour of $a(\phi)$ as $\phi\to\infty$. Here we consider the role of possible subleading corrections. We notice that these corrections do not necessarily go to zero as $\phi\to\infty$ in \eqref{eq:vt_solution}. For example, let us take
\begin{equation}
    \sqrt{1-a(\phi)} = \sqrt{1-a} + \frac{b}{\phi} \, .
\end{equation}
It is clear that $a(\phi)$ asymptotes to $a$ as $\phi\to\infty$, but after doing the integral in \eqref{eq:vt_solution} the correction to the leading behaviour given by the second term behaves as $\log \phi$. Indeed, ignoring constant prefactors we get
\begin{equation}
    V_t \sim \phi^{2 \sqrt{\frac{d-1}{d-2}}\,b} e^{\delta\, \phi} \, ,
\end{equation}
with $\delta$ defined in \eqref{eq:gamma}. Comparing with \eqref{eq:sol-Vt} we see that we can describe this example with our leading order analysis if we allow for $c \sim \phi^{2 \sqrt{\frac{d-1}{d-2}}\,b}$. Notice that the example in section \ref{sec:sugimoto} precisely realise this behaviour (see equation \eqref{eq:DM-Vt}).

As a general lesson, we can include these kind of corrections that do not vanish in the $\phi\to\infty$ limit by promoting $c$ from just a constant to a $\phi$-dependent quantity that may hide subleading corrections. In this way, it may happen that $c\to\infty$ as $\phi\to\infty$ as long as it blows-up slower than an exponential (otherwise it would not represent a subleading behaviour).  

This remark is specially interesting in the $a(\phi)\to0$ case. From \eqref{eq:sol-V} we would conclude that $V\to 0$ if $c$ is a finite constant. However, if allowing $c\to\infty$ because of possible subleading terms, it can happen that $a$ times $c$ remains finite in the $\phi\to\infty$ limit. In this way, we describe a solution in which $\phi'^2\gg V$ (i.e., $a(\phi)\to 0$) without requiring that $V$ vanishes asymptotically.

\newpage
\bibliographystyle{JHEP}
\bibliography{mybib}

\end{document}